\documentclass[11pt, a4paper]{article}
\usepackage{jheppub}
\usepackage[utf8]{inputenc}
\usepackage{lipsum, lmodern}
\usepackage{amsmath, amsthm, amssymb, amsfonts, braket, enumerate, slashed, upgreek, tensor, mathtools, enumitem, bbold}
\usepackage[most]{tcolorbox}
\usepackage[T1]{fontenc}
\usepackage{wrapfig, graphicx, arydshln, float}
\usepackage{hyperref}
\usepackage{microtype}
\graphicspath{ {./figures} }

\allowdisplaybreaks

\newcommand{\pa}{\partial}

\newcommand{\eps}{\epsilon}

\newcommand{\dl}{\delta}
\newcommand{\nb}{\nabla}
\newcommand{\lie}{\mathsterling}

\newcommand{\mc}{\mathcal}

\newcommand{\0}{{(0)}}
\newcommand{\1}{{(1)}}
\newcommand{\2}{{(2)}}

\newcommand{\OP}[1]{\begin{color}{red}(OP: #1)\end{color}}
\newcommand{\cw}{\curlywedge}
\newcommand{\EH}{\mathrm{EH}}
\newcommand{\mat}{\mathrm{matter}}

\newcommand{\su}{\mathsf{u}}
\newcommand{\hpsi}{\widehat{\psi}}
\newcommand{\ba}{\mathsf{a}}
\newcommand{\bF}{\mathsf{f}}
\newcommand{\bbA}{\mathsf{A}}

\newcommand{\rbar}{\overline{r}}
\newcommand{\bbF}{\mathsf{F}}
\title{Covariant phase space and the semi-classical Einstein equation}
\author[a]{Abhirup Bhattacharya,}
\author[a]{Onkar Parrikar}
\affiliation[a]{Department of Theoretical Physics, Tata Institute of Fundamental Research,\\ 1 Homi Bhabha road, Mumbai 400005, India}
\emailAdd{abhirup.bhattacharya@tifr.res.in}
\emailAdd{parrikar@theory.tifr.res.in}

\abstract{The covariant phase space formalism in general relativity is a covariant method for constructing the symplectic two-form, Hamiltonian and other conserved charges on the phase space of solutions to the Einstein equation with classical matter. In this note, we consider a generalization of this formalism to the semi-classical Einstein equation coupled to quantum matter. Given a family of solutions in semi-classical gravity, we define the semi-classical symplectic two-form -- a natural generalization of the classical sympelctic two-form -- as the sum of the gravitational symplectic form and the Berry curvature associated to the quantum state of matter. We show that the semi-classical symplectic two-form is independent of the Cauchy slice, and satisfies the quantum generalization of the classical Hollands-Iyer-Wald identity. For small perturbations, we also extend our discussion to gauge-invariantly defined subregions of spacetime, where the quantum contribution is replaced by the Berry curvature of certain special purifications involving the Connes cocycle. In the AdS/CFT context, the semi-classical symplectic form defined here is naturally dual to the Berry curvature in the boundary CFT. }

\begin{document}

\maketitle

\section{Introduction} \label{sec:intro}
The covariant phase space formalism \cite{crnkovic1987covariant} provides a manifestly covariant way (\emph{i.e.}, without picking a time coordinate) to think about the classical phase space of solutions to the Einstein equation with its associated symplectic structure, Hamiltonian and other conserved charges. The formalism has found many important applications -- two significant examples include Iyer and Wald's generalization of the the Bekenstein-Hawking area formula for black hole entropy to higher derivative theories of gravity \cite{Wald:1993nt, Iyer:1994ys, Iyer:1995kg}, and Wald and Zoupas' general formulation of the notion of conserved charges in general relativity \cite{Wald:1999wa}. More recent developments have used the formalism to study the stability of black hole solutions under perturbations \cite{Hollands:2012sf}, and to propose new definitions of dynamical black hole entropy \cite{Hollands:2024vbe, Wall:2015raa}. The formalism has also found a wide variety of applications in the context of the AdS/CFT correspondence, such as the proof of the equality of bulk and boundary relative entropies \cite{Jafferis:2015del}, derivations of Einstein's equation from entanglement in the boundary CFT \cite{Faulkner:2013ica, Faulkner:2017tkh, Haehl:2017sot, Dong:2017xht, Lewkowycz:2018sgn, Haehl:2019fjz}, proofs of gravitational positive energy theorems from CFT entropic inequalities \cite{Lashkari:2016idm}, and the derivation of the bulk dual of boundary Berry curvature at large-$N$ \cite{Belin:2018fxe, Belin:2018bpg}. Beyond AdS/CFT, the covariant phase space formalism has been used to investigate near horizon symmetries of black holes \cite{Flanagan:2015pxa, Hawking:2016msc, Hawking:2016sgy, Haco:2018ske}, asymptotic symmetries and surface charges in general relativity \cite{Chandrasekaran:2021hxc, Chandrasekaran:2021vyu, Ciambelli:2021nmv, Shi:2020csw, Ciambelli:2022cfr, Ciambelli:2022vot}, soft theorems \cite{He:2020ifr, He:2023bvv}, and entanglement edge modes in gauge theories \cite{Donnelly:2016auv, Balasubramanian:2018axm}. 

Many of the results mentioned above work in the context of classical general relativity coupled to classical matter. In particular, the matter contributes to the symplectic two-form on phase space, and the total (gravity plus matter) symplectic form can be defined without choice of a Cauchy surface. It is natural to wonder whether this formalism finds a natural generalization to semi-classical gravity. In semi-classical gravity, one treats matter quantum mechanically and finds the metric by solving the semi-classical Einstein equation:
\begin{equation}
    R_{ab}- \frac{1}{2}R\,g_{ab}+ \Lambda\,g_{ab} = 8\pi G_N\langle \psi|T_{ab}|\psi\rangle,
\end{equation}
where the right hand side involves the expectation value of the matter stress tensor. It is usually inconsistent to treat a physical system part classically and part quantum mechanically, so what is really being said here is that to a good approximation in the $G_N \to 0$ limit, the metric can be taken to be in a coherent state, while the matter can be in some more general quantum state. In order to have significant backreaction in this limit, a standard trick \cite{Flanagan:1996gw} is to introduce a large matter central charge $c$, so that back-reaction is controlled by the parameter $\epsilon = c\,G_N$. The semi-classical approximation has found a large number of applications; some recent applications include Page curve calculations for evaporating black holes \cite{Fiola:1994ir, Engelhardt:2014gca, Almheiri:2019psf, Penington:2019npb}, constructing endpoints of the super-radiant instabilities of rotating and charged black holes \cite{Kim:2023sig, Choi:2024xnv}, bulk reconstruction \cite{Levine:2020upy, Soni:2024oim}, singularity theorems \cite{Bousso:2025xyc} etc. 

The purpose of this note is to generalize the covariant phase space formalism to semi-classical gravity. We imagine that we are given a family $\mathcal{P}_{\text{q}}$ of solutions $(g_{ab}(\lambda^i),\psi(\lambda^i))$ to the semi-classical Einstein equation, where $\lambda^i$ are some parameters, which we can think of as coordinates on the space of these solutions. A concrete setting to have in mind is asymptotically AdS spacetimes, where the role of the parameters $\lambda^i$ is played by the sources for various single/multi-trace operators in the boundary CFT; this is the setting we will focus on through most of this paper. We propose that the space $\mathcal{P}_{\text{q}}$ also admits a natural, closed two form, which we call the \emph{semi-classical symplectic form}, given by
\begin{equation}
\bbF = \Omega_{\text{grav.}}+ \bF,
\end{equation}
where $\Omega_{\text{grav.}}$ is the symplectic 2-form associated with classical metric deformations, while $\bF$ is the \emph{Berry curvature} associated with the quantum matter:
\begin{equation}
    \bF = \delta \ba,\;\;\; \ba= -i\langle \psi| \delta \psi\rangle,
\end{equation}
where $\delta = \sum_i d\lambda^i\frac{\partial}{\partial \lambda^i}$ is the exterior derivative on $\mathcal{P}_{\text{q}}$. We will show that the semi-classical symplectic form $\bbF$ defined in this way is indepdendent of the choice of Cauchy slice, and indeed reduces to the classical matter symplectic form when the matter is taken to be in a coherent state. Furthermore, given some asymptotically Killing vector field $\zeta$, we show that the semi-classical symplectic form is invariant under the corresponding diffeomorphism. We show how to construct a corresponding charge $H_{\zeta}$ (including classical plus the appropriate quantum corrections coming from matter) which satisfies 
\begin{equation}
    - I_{V_{\zeta}}\bbF = \delta H_{\zeta}.
\end{equation}
This is essentially the statement that the charge $H_{\zeta}$ generates the action of the diffeomorphism $\zeta$ on phase space. This identity is sometimes referred to as the Hollands-Iyer-Wald identity in the classical gravity context, and here we generalize it to semi-classical gravity. For small perturbations around a background metric and state, we also extend our discussion to the case of gauge-invariantly defined subregions in gravity, where the role of the matter contribution is played by the Berry curvature of a certain purification of the subregion state involving the Connes cocycle. 

In the AdS/CFT context, it was argued in \cite{Belin:2018fxe, Belin:2018bpg} that in the large-$N$ limit, if we take the parameters $\lambda^i$ to be $O(N^2)$ classical sources for single-trace operators, then the Berry curvature of the boundary CFT is dual to the symplectic form of the corresponding bulk matter fields. It is natural to try to extend this to more general sources, including $O(1)$ sources for multi-trace operators. In this context, the bulk is more appropriately  described within the realm of semi-classical gravity with quantum matter. We argue that our semi-classical symplectic form is the natural bulk dual to the boundary Berry curvature within the semi-classical gravity approximation, i.e., not including corrections from quantum gravity and non-perturbative effects. Thus, the semi-classical symplectic form that we define fits naturally within the AdS/CFT dictionary.  

The rest of the paper is organized as follows: in section \ref{sec:prelim}, we review the covariant phase space formalism for gravity coupled to classical matter. In section \ref{sec:cov-quant}, we construct and study the semi-classical symplectic form for complete Cauchy slices. In section \ref{sec:quant-HW-sub}, we extend this discussion to the case of subregions of the bulk spacetime. In section \ref{sec:Ads/cft}, we discuss the duality between the semi-classical symplectic form in the bulk and the Berry curvature in the boundary CFT.

\section{Preliminaries} \label{sec:prelim}
In this section, we will review the background material needed for extending the covariant phase space formalism of gravity to include quantum matter.
\subsection{A primer on field spaces and infinite dimensional manifolds}\label{sec:primer}
\begin{figure}
    \centering
    \includegraphics[height=8cm]{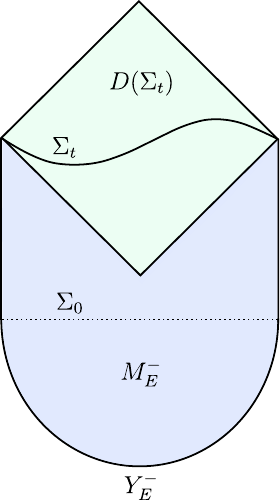}
    \caption{We consider spacetimes obtained by specifying asymptotic boundary conditions on the boundary $Y^-$, which involves a Euclidean part and a Lorentzian part. This data completely specifies the semi-classical geometry within the diamond $D(\Sigma_t)$ defined at asymptotic time $t$. }
    \label{fig:setup}
\end{figure}

Let $M$ be a $D = (d+1)$-dimensional spacetime manifold, and let $\mc{C}$ denote the set of field configurations on $M$ corresponding to all the dynamical fields in the theory. Following \cite{Harlow:2019yfa}, we will refer to $\mc{C}$ as the \emph{configuration space} of the theory. In this paper, we will mostly be interested in asymptotically AdS spacetime manifolds which have both Euclidean and Lorentzian (timelike) asymptotic boundaries (for instance, see figure \ref{fig:setup}), but many aspects of our analysis are more general. 
Let $\left\{\phi^a\right\}$ denote the set of all fundamental fields of the theory and let 
$$L[\phi^a]  = f[\phi^a]\,\epsilon_M,$$ 
be a local Lagrangian density, thought of as a top-degree form on $M$ with $\epsilon_M$ being the volume form. Then, varying the Lagrangian with respect to $\phi^a$, we get 
\begin{equation} \label{eq:delL}
    \delta L = E_a \delta\phi^a + d\theta[\delta \phi^a],
\end{equation}
where the sum over $a$ is assumed. Here $E_a$ is a top-degree form proportional to the equation of motion and $d\theta$ is a total derivative term. Since we require the action to be stationary, we must have 
\begin{equation}
    \delta S = \int_M E_a \delta\phi^a + \int_{\partial M} \theta[\dl \phi^a] = 0. 
\end{equation}
In order to have a well-defined variational problem, we need to impose boundary conditions on the fields so that $\theta|_{\partial M}$ vanishes. In cases where standard normalizable boundary conditions on $\phi^a$ are not enough, we typically add a boundary term $\ell[\phi^a] \epsilon_{\partial M}$ to the Lagrangian density and impose boundary conditions such that $(\theta + \delta (\ell \epsilon_{\partial M}))|_{\partial M} = 0$. Since $\ell$ can be an arbitrary local functional, this buys us enough freedom to ensure the validity of the variational principle. A well-known example of such a boundary term is the Gibbons-Hawking-York term of Einstein gravity \cite{Gibbons:1976ue, York:1972sj}. 

Once we have a well-defined variational problem, we define the \emph{covariant phase space} $\mc{P} \subset \mc{C}$ as the set of all field configurations that satisfy $E_a = 0$. In situations where the initial value problem is well-defined, $\mc{P}$ gives a covariant way of defining the canonical phase space of the theory, since the space of initial data on a Cauchy surface $\Sigma$ must be in one-to-one correspondence with the set $\mc{P}$ of solutions to equations of motion.\footnote{To ensure that the initial value problem is well-defined, it is usually assumed that the spacetime is globally hyperbolic. Note, however, that AdS does not admit a complete Cauchy surface, and is therefore not globally hyperbolic. Nevertheless, one can obtain a well-defined phase space for theories in AdS by specifying suitable boundary conditions at infinity. See for instance \cite{Breitenlohner:1982bm, Wald:1980jn,Ishibashi:2003jd,Ishibashi:2004wx}. In this paper we will always assume that such boundary conditions have been specified.} In gauge theories, one must in addition quotient $\mc{P}$ by the group of gauge transformations in order to obtain a well-defined (i.e., non-degenerate) symplectic form on this phase space. We will temporarily ignore this subtlety and return to issues of gauge invariance in section \ref{sec:class-cov}.

A standard technical assumption in the covariant phase space formalism is that $\mc{C}$ and $\mc{P}$ are manifolds with enough structure, so that standard techniques of differential calculus can be applied \cite{Harlow:2019yfa, crnkovic1987covariant}. For instance, it is assumed that $\mc{C}$ has a tangent bundle $T\mc{C}$ whose sections define vector fields, and that for every such smooth vector field $V$ on $\mc{C}$, there exists a Lie derivative $\mc{L}_V$ on $\mc{C}$. Moreover, one also demands that there exists an exterior derivative $\dl$ on $\mc{C}$ satisfying $\dl^2 = 0$ such that finite degree differential forms are well-defined. Let us denote interior multiplication on $\mc{C}$ with respect to a vector field $V$ as $I_V$. Then $\lie_V$, $\dl$ and $I_V$ are required to satisfy Cartan's magic formula
\begin{equation}
    \lie_V = I_V \dl + \dl I_V.
\end{equation}
It is also assumed that the covariant phase space $\mc{P}$ is a submanifold of $\mc{C}$, and that one can define differential forms on $\mc{P}$ by pulling back from the ambient space $\mc{C}$.

A pedestrian way to think about differential forms and exterior derivatives on $\mc{P}$ is 
to imagine $\mc{P}$ to be an $n$-parameter family of solutions $\phi^a(\lambda_1, \lambda_2, \dots,\lambda_n)$. For instance, the $\{\lambda_i\}$ could be some parameters in the specified sources at asymptotic infinity in AdS. From this point of view, the exterior derivative is defined as $\delta = \sum_i d\lambda_i\frac{\pa}{\pa \lambda_i}$, and the statement $\delta^2 = 0$ follows trivially from the commutativity of the partial derivatives $\partial/\partial \lambda_i$ and $\partial/\partial\lambda_j$. We will often use the notation $\left\{\frac{\delta}{\delta \phi^a(x)}\right\}$ and $\left\{\delta \phi^a(x)\right\}$ to denote a local coordinate basis for the tangent space and the cotangent space of $\mc{C}$ respectively, at some point $\phi^a(x)\in \mc{C}$. In this notation, the exterior derivative takes the form 
\begin{equation}
\delta = \int_M \delta\phi^a(x) \frac{\delta}{\delta \phi^a(x)}.
\end{equation}

One thing to keep in mind is that we will often need to talk about configuration space differential forms valued in spacetime differential forms. We say that $\alpha$ is a form of degree $(p,n)$ if it is an $n$-form on configuration space which takes as input $n$ vector fields on $\mathcal{C}$ and gives a $p$-form on spacetime. Thus, for any point $x\in M$ and $\phi^a \in \mathcal{C}$,  
\begin{equation}
    \alpha|_{x,\phi^a}: \otimes^n T_{\phi}\mathcal{C} \to \wedge^p T^*_xM.
\end{equation}
From this point of view, it seems natural to follow the convention that the spacetime exterior derivative $d$ commutes with $\delta$:\footnote{In the literature, it is occasionally assumed that $\delta$ and $d$ \emph{anti}-commute (for instance, see Appendix A of \cite{Barnich:2007bf}). This assumption lets one organize the two exterior derivatives $\delta$ and $d$ into a differential \emph{bi}-complex that extends the ordinary de Rham complex of $M$. As a result one can meaningfully talk about vertical and horizontal cohomologies in the space of fields. In this paper we will not use this language and therefore the commutativity convention suffices.}
\begin{equation}
    \left[d,\delta\right]=0.
\end{equation}

In this paper, we will denote the wedge product between differential forms on $\mathcal{C}$ by $\curlywedge$, while the symbol $\wedge$ will be used to denote the wedge product for spacetime differential forms. 

As noted before, we can think of the phase space $\mc{P}$ as the submanifold of $\mc{C}$ picked out by the equations of motion $E_a[\phi]=0$. The tangent space of $\mathcal{P}$ at some point $\phi^a$ corresponds to the set of linearized solutions to the equations of motion around the background solution $\phi^a$. Vector fields on $\mathcal{P}$ are correspondingly sections of the tangent bundle of $\mathcal{P}$. Finally, differential forms on $\mc{C}$ can be pulled back in the standard way to differential forms on $\mathcal{P}$. We will use the notation $F_*$ to denote pullbacks under the embedding $F: \mc{P} \to \mc{C}.$

\subsection{The covariant symplectic form} \label{sec:class-cov}
We now turn to the definition of the symplectic form on $\mathcal{P}$, which is a closed, non-degenerate $(D-1, 2)$-form on $\mathcal{P}$. We define the symplectic current $\omega$ as \cite{Lee:1990nz}
\begin{equation} \label{eq:SympCurrent}
    \omega = F_*(\delta \theta),
\end{equation}
where $\theta$ is the $(D-1,1)$-form defined in eq. \eqref{eq:delL}, and $F_*$ is the pullback from $\mc{C}$ to $\mc{P}$. Let $\Sigma$ be a Cauchy slice in spacetime. Then the symplectic form $\Omega_{\Sigma}$ is defined as
\begin{equation}
    \Omega_\Sigma = \int_\Sigma \omega.
\end{equation}
Clearly, $\Omega_\Sigma$ is closed since $\delta \theta$ is an exact form on $\mc{C}$ and the pullback of an exact form is also exact. Moreover, it is independent of the choice of Cauchy slice $\Sigma$. To see this consider
\begin{align}
    \begin{split}
        d\omega &= F_*(\delta d\theta) \\
        &= F_*(\delta (\delta L - E_a \delta\phi^a)) \\
        &= - F_*(\delta E_a \curlywedge \delta \phi^a),
    \end{split}
\end{align}
where in the first line we have used the fact that $d$ commutes with $\delta$ and $F_*$, in the second line we have used eq. \eqref{eq:delL}, and in the third line we have used $\dl^2 = 0$. Since the pullback of $\delta E_a$ to $\mc{P}$ vanishes (since $\mc{P}$ by definition is the space of solutions to equations of motion), we see that $d \omega = 0$ and therefore $\Omega_\Sigma$ is independent of the choice of $\Sigma$.\footnote{Here we have assumed that the boundary conditions at asymptotic infinity are such that there is no symplectic flux at infinity.}

Although, $\Omega_\Sigma$ is closed, it is degenerate if the theory under consideration has gauge symmetries \cite{Lee:1990nz, Donnelly:2016auv}. 
In order to obtain a non-degenerate symplectic form, one quotients $\mathcal{P}$ with the group $G$ of ``small'' gauge transformations which decay sufficiently fast at asymptotic infinity. In the literature, $\mc{P}$ is often referred to as the pre-phase space whereas $\mc{P}/G$ is the true phase space. This distinction will not be too important for the purposes of this paper. 


As a final comment, note that the definition of $\theta$ in eq. \eqref{eq:delL} is not unique, and this may lead to ambiguities in the definition of the covariant symplectic form. There are two potential sources of ambiguity:
\begin{enumerate}
    \item We can always add a boundary term of the form $\dl (\ell\, \epsilon_{\partial M})$ to the Lagrangian, where $\epsilon_{\pa M}$ is some extension of the boundary volume form to the bulk. This modifies $\theta$ to
    \begin{equation}
        \theta \to \theta + \dl (\ell \epsilon_{\partial M}).
    \end{equation}
    Note that this does not lead to an ambiguity in the definition of the symplectic current since the extra term is $\dl$-exact.
    \item We can always add to $\theta$ a $(D-2, 1)$-form that is $d$-exact. In other words, modifying $\theta$ by
    \begin{equation}
        \theta \to \theta + d\beta
    \end{equation}
    does not change equation \eqref{eq:delL}. However, the symplectic form picks up an additional boundary term:
    \begin{equation}
        \Omega_{\Sigma} \to \Omega_{\Sigma} + \int_{\partial \Sigma} \delta \beta.
    \end{equation}
    This ambiguity is harder to resolve. In certain cases asymptotic boundary conditions on the fields can ensure that $\delta \beta|_{\partial \Sigma}$ vanishes. But this is not guaranteed in general. A proposal for fixing this ambiguity for global states in AdS was given in \cite{Harlow:2019yfa}. We will not have anything further to add to this discussion; we will simply take $\beta=0$ as a minimal prescription. 
\end{enumerate}

\subsection{Gravitational phase space with classical matter} \label{sec:class-grav}
Let $M$ be a $(d+1)$-dimensional spacetime and let 
\begin{equation}
    L[g,\phi^A] = L_{\EH}[g] + L_{\mat}[g, \phi^A] + d\ell[g],
\end{equation}
be the Lagrangian density describing Einstein gravity with a cosmological constant coupled to matter, where $g=g_{ab}dx^{a}dx^{b}$ is the spacetime metric, $\{\phi^A\}$ collectively denotes all the matter fields,
\begin{align}
    L_{\text{EH}} &= \frac{1}{16 \pi G_N} \left( R - 2 \Lambda \right) \epsilon_M,
\end{align}
and $L_{\text{matter}}$ is the matter Lagrangian. Further, $\ell$ is a $(D-1,0)$ form (covariantly constructed from the metric, Riemann tensor and its derivatives) such that its restriction to the boundary reproduces the Gibbons-Hawking term, plus any potential counterterms: 
\begin{equation} \label{eq:GHY}
\ell|_{\pa M} = \frac{1}{8\pi G_N} K \epsilon_{\partial M} + \ell_{\text{ct}}.
\end{equation} 
The Gibbons-Hawking-York term ensures that the variational principle is well-defined with Dirichlet boundary conditions at asymptotic infinity, while the counterterms are added in order to cancel any divergences in holographic renormalization. 
Varying the Lagrangian with respect to the metric $ g$ and the matter fields $\phi^A$, we get
\begin{equation}\label{eq:Lvar}
    \dl L = E_{ab}\dl g^{ab} + E_{A} \dl \phi^A +  d\theta^g + d\theta^\phi,
\end{equation}
Here $E_{ab}$ and $E_{A}$ are proportional to the equations of motion for the metric and the matter fields respectively, and $d\theta^g$ and $d\theta^\phi$ are the corresponding total derivative terms. For instance, if we take the matter sector to consist of a free, real, massive scalar field minimally coupled to gravity, then we get:
\begin{align} \label{eq:LvarTerms}
    \begin{split}
        E_{ab} &= \frac{1}{16\pi G_N}\left(R_{ab} - \frac{1}{2}g_{ab}R + \Lambda g_{ab} - 8\pi G_N T_{ab}^{\mat} \right) \eps_M, \\
        T_{ab}^{\text{matter}} &= \nabla_a \phi \nabla_b \phi - g_{ab} \mc{L}_{\text{matter}},\\
        E_\phi &= \nabla_a \nabla^a \phi - m^2 \phi, \\
        \theta^g|_{(g,\phi)} &= \frac{1}{16\pi G_N} \left( \nabla_a \dl g^{ab} - \nabla^b \dl g \indices{^a_a}\right) \eps_{b} + \delta \ell, \\  
        \theta^\phi|_{(g,\phi)} &= - (\dl \phi \nabla^b \phi)\, \eps_{b},
    \end{split}
\end{align}
where the covariant derivative $\nabla_a$ is with respect to the background metric around which the variation is taken, and
$\epsilon_b = \sqrt{-g}\, \epsilon_{ba_2\cdots a_D} dx^2 \wedge  \cdots \wedge dx^D. $
The corresponding gravitational and matter symplectic forms are given by\footnote{Note from equation \eqref{eq:SympCurrent} that the symplectic from is the pullback of $\dl \theta$ to $\mc{P}$. Henceforth, we will not explicitly write the pullback when we are talking about phase space.}
\begin{align}
    \omega_{\text{matter}}|_{(g,\phi)} &=  - 2 (\dl \phi \cw \nabla^a \dl \phi)\,\eps_a - 2 \nabla_a \phi (\dl \phi \cw \dl g^{ab})\, \epsilon_b + g_{cd} \nabla^a \phi (\dl \phi \cw \dl g^{cd})\, \epsilon_a , \\
    \omega_{\text{grav}}|_{(g,\phi)} &= \frac{1}{8\pi G_N} P^{abcdef}(\dl g_{bc} \cw \nb_d \dl g_{ef} )\,\eps_a,\\
    P^{abcdef} &= g^{ae}g^{bf}g^{cd} - \frac{1}{2}g^{ad}g^{be}g^{cf} - \frac{1}{2} g^{ab}g^{cd}g^{ef} - \frac{1}{2} g^{ae}g^{bc}g^{fd} + \frac{1}{2} g^{ad}g^{bc}g^{ef}.
\end{align}

Let $\zeta$ be a vector field on $M$. To every such vector field we can define a corresponding vector field on $\mathcal{C}$ as
\begin{equation} \label{eq:CSVF}
    V_\zeta = \int_M dx\, \left( \lie_\zeta g_{ab}(x) \frac{\dl}{\dl g_{ab}(x)} + \lie_\zeta \phi^A(x) \frac{\dl}{\dl \phi^A(x)}\right).
\end{equation}
Note that since the coefficients of this vector field correspond to a pure diffeomorphism, this vector field naturally restricts to a vector field on $\mc{P}$, when evaluated on $\mathcal{P}$. Even if we only consider some $n$-parameter family of solutions, we will always assume that this space has been suitably extended such that the action of pure diffeomorphisms leaves us within $\mc{P}$. Let $\lie_{V_\zeta}$ be the Lie derivative with respect to the vector field $V_{\zeta}$ and $\eta$ be a differential form that is locally constructed out of the fields. Then $\lie_{V_\zeta} \eta$ can be interpreted as computing the variation of $\eta$ under a spacetime diffeomorphism. Now, recall that the Lie derivative $\lie_{V_\zeta}$ satisfies Cartan's magic formula:
\begin{equation}
    \lie_{V_\zeta} \eta = \delta(I_{V_\zeta} \eta) + I_{V_\zeta} (\dl \eta).
\end{equation}
We say that the configuration space form $\eta$ is \emph{covariant} under the diffeomorphism generated by $\zeta$ if $\lie_{V_\zeta} \eta = \lie_\zeta \eta$, where the right hand side is the ordinary spacetime Lie derivative in $M$. It is important to note that the configuration space Lie derivative only acts on the \emph{dynamical} fields of the theory whereas the spacetime Lie derivative acts on all fields, including any potential external sources. We say that a theory is covariant with respect to a diffeomorphism $\zeta$ if the following holds:
\begin{equation}
    \lie_{V_\zeta} L = \lie_\zeta L.
\end{equation}
A gravitational theory should be covariant under all diffeomorphisms. In fact, it can be shown that any theory whose Lagrangian satisfies the following two conditions must be diffeomorphism covariant \cite{Iyer:1994ys}:
\begin{enumerate}
    \item $L$ must not contain any external/background fields,
    \item $L$ can only be a function of arbitrary covariant derivatives of the Riemann tensor $\nabla_{a_1} \nabla_{a_2} \cdots \nabla_{a_n} R_{abcd}$ and its traces, and arbitrary covariant derivatives of the matter fields $\nabla_{a_1} \nabla_{a_2} \cdots \nabla_{a_n} \phi^A$ and its traces, for any $n \geq 0$.
\end{enumerate}
This set of conditions is sufficiently general to accommodate all higher derivative theories of gravity, including Einstein gravity minimally coupled to matter. Note that this is simply a re-formulation of the familiar statement that all diffeomorphisms are gauge symmetries of gravity.

\subsection{Boundary charges for diffeomorphisms} \label{sec:boundary-charges}
Let $\zeta$ be a vector field on $M$ that generates a one-parameter family of diffeomorphisms. We define the bulk current associated to $\zeta$ as the $(D-1,0)$ form \cite{Wald:1999wa} as 
\begin{equation}\label{eq:Jdef}
    J_{\zeta} = I_{V_\zeta}\theta - i_\zeta L,
\end{equation}
where $L = L_{\EH} + L_{\mat} + d\ell$, $\theta = \theta^g + \theta^\phi$, and recall that $\ell$ is an arbitrary extension of the boundary Gibbons-Hawking term (plus counterterms) to the bulk. 
Naively it may seem like $J_{\zeta}$ depends on this arbitrary choice. However, by covariance, the $\ell$ dependence of $J_{\zeta}$ reduces to a boundary term:
\begin{eqnarray}
    J_{\zeta}\Big|_{\ell-\text{dep.\;terms}} &=& I_{V_{\zeta}}\delta \ell - i_{\zeta}d\ell\nonumber\\
    &=& \lie_{V_{\zeta}}\ell - i_{\zeta}d\ell\nonumber\\
    &=& \lie_{\zeta}\ell - i_{\zeta}d\ell\nonumber\\
    &=& d\left(i_{\zeta}\ell\right),
\end{eqnarray}
so the ambiguity of extending $\ell$ into the bulk drops out once we integrate the current over a complete Cauchy surface. If however the integral is over a subregion, say, the homology surface of an entanglement wedge in AdS/CFT, then one may pick up an additional ambiguous contribution from the extremal surface. If $\zeta$ vanishes on the extremal surface, then this ambiguity drops out.

In order to show that $J_\zeta$ conserved on-shell, we act on both sides of equation \eqref{eq:Jdef} with $d$\footnote{The standard convention is to define the current $\widetilde{J}$ as a spacetime 1-form, and the conservation equation is given by $d\star \widetilde{J}=0$. Following \cite{Wald:1999wa}, we define the current as the Hodge dual of the standard definition -- $J = \star \widetilde{J}$ -- and so is a spacetime $(D-1)$ form. The corresponding conservation equation is simply $dJ=0$.  } :
\begin{align}
    \begin{split} \label{eq:dJ}
        dJ_\zeta &= d \left( I_{V_\zeta} \theta - i_\zeta L \right) \\
        &= I_{V_\zeta} \left( \dl L - E_{ab} \dl g^{ab} - E_A \dl \phi^A \right) - d\left(i_\zeta L\right) \\
        &= \lie_{\zeta} L  - d\left(i_\zeta L\right) - I_{V_{\zeta}}\left( E_{ab} \dl g^{ab} + E_A \dl \phi^A \right) \\
        &= - E_{ab} \,\lie_\zeta g^{ab} - E_A\, \lie_\zeta \phi^A \\
        &= 0,
    \end{split}
\end{align}
where in the second line we have used equation \eqref{eq:delL}, in the third line we have used covariance to write $I_{V_\zeta} \dl L = \lie_{\zeta} L$, and in the fourth line we have used Cartan's magic formula. Let $\Sigma$ be a Cauchy slice. The charge associated to the vector field $\zeta$ is defined as
\begin{equation} \label{eq:WZCurrent}
    H_{\Sigma}(\zeta) = \int_{\Sigma} J_{\zeta}.
\end{equation}
Since $J_\zeta$ is closed, it follows that $H_{\Sigma}(\zeta)$ is independent of $\Sigma$ if $\zeta$ generates a diffeomorphism that decays sufficiently fast at asymptotic infinity. In fact, as we will see below, in this case the charge identically vanishes. However, if $\zeta$ is a large diffeomorphism (i.e., it acts on the asymptotic boundary), then the charge is non-trivial, but may not be conserved in general. In addition, if $\zeta$ asymptotically approaches a Killing vector field of the background metric $g$, we get a non-trivial, conserved charge.\footnote{Since $\zeta$ is only required to be an asymptotic Killing vector field, the corresponding charge can be defined for an open subset in phase space, corresponding to metrics with fixed asymptotic structures.}

It can be shown that in any diffeomorphism covariant theory of gravity, there exists a $(D-2, 0)$-form $Q_\zeta$ and a $(D-1, 0)$-form $C_\zeta$ such that
\begin{equation} \label{eq:charge-os}
    H_{\Sigma}(\zeta) = \int_{\partial \Sigma} Q_\zeta + \int_{\Sigma} C_{\zeta},
\end{equation}
where $C_{\zeta}$ vanishes on-shell \cite{Iyer:1994ys, Iyer:1995kg}. We will not prove this statement in full generality here (see \cite{Iyer:1994ys, Iyer:1995kg} for more details). Instead, we will simply note that it holds in Einstein gravity with matter, where one can show by explicit computation that (see Appendix \ref{app:deriv} for the derivation):
\begin{align} \label{eq:Charge-EH}
    Q_\zeta &= - \frac{1}{16\pi G_N} \epsilon^{ab} \nabla_a \zeta_b + i_\zeta \ell, \\
    C_{\zeta} &= \frac{1}{8 \pi G_N} \epsilon_a \left( R^{ab} - \frac{1}{2} g^{ab} R + \Lambda g^{ab} - 8\pi G_N T^{ab}_{\text{matter}} \right) \zeta_b.
\end{align}
Note that the $C_{\zeta}$ term is proportional to the Einstein equation and so drops out once we pull back to the covariant phase space $\mc{P}$. From here, it is immediately clear that for any $\zeta$ which vanishes sufficiently fast asymptotically, the charge $H_{\Sigma}(\zeta)$ vanishes on-shell. Also note that the ambiguity in the choice of $\ell$ in the bulk trickles down to equation \eqref{eq:Charge-EH}. As was pointed out before, this is irrelevant if $\zeta$ vanishes on the surface over which $Q_{\zeta}$ is integrated.

In general, $H_{\Sigma}(\zeta)$ will not be zero for large diffeomorphisms, but will nevertheless be a boundary term. For instance, let $\Sigma$ be a Cauchy slice and suppose $\zeta$ asymptotically approaches a Killing vector field for some family of solutions. Then the boundary charge $H_{\Sigma}(\zeta)$ for this family of solutions can be written as a local integral of the extrinsic curvature of $\partial \Sigma$ (see Appendix \ref{app:deriv}):
\begin{equation}\label{eq:BYcharge}
    H_{\Sigma}(\zeta) = - \frac{1}{8\pi G_N} \int_{\partial \Sigma} \epsilon_{a} \left( K^{ab} - h^{ab} K \right)  \zeta_b,
\end{equation}
where $K^{ab}$ is the extrinsic curvature of $\pa \Sigma$ with respect to the (radially) outward pointing normal vector, with $K = h^{ab} K_{ab}$. The quantity in brackets is called the Brown-York stress tensor \cite{Brown:1992br}
\begin{equation}
    T^{ab}_{\text{BY}} = - \frac{1}{8\pi G_N} \left( K^{ab} - h^{ab} K \right),
\end{equation}
and is conserved on account of Einstein's equations.\footnote{In the presence of matter, one also needs to assume that there is no matter flux at infinity. This is certainly the case if reflecting boundary conditions are imposed.} This makes it clear that the charge in equation \eqref{eq:BYcharge} is conserved when $\zeta$ is a Killing vector field in the boundary. 

As a second example, let $r$ be a homology surface in the entanglement wedge associated to the right boundary $R$ of a family of asymptotically AdS black hole solutions, and let $X$ denote the extremal surface corresponding to the right boundary. Then $\partial r = X \cup R$. We will assume that this family of solutions admit an asymptotically timelike Killing vector field $\xi$ with the following behavior on the extremal surface:
\begin{equation}
    \xi\big|_{X} = 0,\quad\nabla_{a}\xi_{b}\big|_{X}= \kappa\, n_{ab},
\end{equation} 
where $n_{ab}$ is the binormal on $X$, and $\kappa$ is a constant called the surface gravity. If we choose the binormal to be normalized as $n_{ab} n^{ab} = -2$, then $\kappa = 2\pi$. Stationary black holes admit such a vector field which is Killing everywhere and satisfies the above properties at the bifurcate Killing horizon. Moreover, the constancy of $\kappa$ on the horizon is the zeroth law of thermodynamics for stationary black holes \cite{wald2010general}.\footnote{Note that in \cite{Jensen:2023yxy} it was argued that the constancy of $\kappa$ on arbitrary gauge invariantly defined surfaces is necessary for the existence of a KMS state whose modular flow is generated by the vector field $\xi$. This is certainly the case for stationary black holes.} In this case, the charge $H_{r}(\xi)$ (defined as  the integral of $J_{\xi}$ over $\Sigma_R$) has two contributions, one from the asymptotic boundary $R$ and one from the extremal surface $X$. From equation \eqref{eq:Charge-EH}, the term localized at the bifurcation point $X$ can be shown to be proportional to $\frac{1}{4G_N}$ times the area of the bifurcation surface $\mc{A}(X)$, and the term localized at the asymptotic boundary is a local integral of the Brown-York stress tensor $T^{ab}_{\text{BY}}$ (see Appendix \ref{app:deriv}):
\begin{equation} \label{eq:bdy-charge-wedge}
    H_{\Sigma}(\xi) = \int_R \epsilon_a T^{ab}_{\text{BY}} \xi_b - \frac{\mc{A}(X)}{4 G_N} ,
\end{equation}
where in the first term on the right hand side, the integration is over a boundary Cauchy surface in $R$.  By the extrapolate dictionary of AdS/CFT, the Brown-York stress tensor can be identified with the CFT stress tensor \cite{Balasubramanian:1999re}. Therefore the asymptotic boundary term in equation \eqref{eq:bdy-charge-wedge} is precisely the CFT modular Hamiltonian for the thermofield double state associated to the right boundary. In fact, the same conclusion holds for wedge like regions in any static spacetime with a bifurcate Killing horizon. 

\subsection{The classical Hollands-Wald identity} \label{sec:class-HW}
The Hollands-Wald (HW) identity is a general identity that shows that the charges defined above actually generate the action of the large diffeomorphisms on phase space \cite{Hollands:2012sf}:
\begin{equation}
    -I_{V_{\zeta}}\Omega_{\Sigma} = \delta H_{\Sigma}(\zeta).
\end{equation}
This identity has been incredibly useful in various contexts. For instance, in \cite{Hollands:2012sf}, this identity was used to derive a criterion for the stability of black holes in terms of the canonical energy. In the AdS/CFT context, the HW identity was used in \cite{Lashkari:2015hha} to derive a perturbative bulk formula for the boundary relative entropy, and to relate the positivity of bulk canonical energy to positivity of boundary relative entropy. This identity was also used in \cite{Faulkner:2013ica} to derive the linearized Einstein equation around vacuum AdS assuming the Ryu-Takayanagi formula and the CFT first law of entanglement. This result was then generalized to second order in \cite{Faulkner:2017tkh} and to more general backgrounds in \cite{Lewkowycz:2018sgn}, again making crucial use of the HW identity.  In this section, we will review the proof of this identity for gravity with classical matter. The goal of this paper is to generalize this identity to the case of semi-classical gravity with quantum matter. In a companion paper \cite{upcoming}, we will use these results to extend the work on deriving Einstein's equations from entanglement to second order, including quantum corrections in the bulk.  

The first step in the proof is to take a variation of $J_{\zeta}$, which yields the following $(D-1, 1)$ form:
\begin{align}
    \begin{split}
        \dl J_\zeta &= \dl (I_{V_\zeta} \theta) - \dl(i_\zeta L) \\
        &= \lie_{V_\zeta} \theta - I_{V_\zeta} \dl \theta - i_{\zeta} \dl L,
    \end{split}
\end{align}
where, as before, we have defined $L = L_{\EH} + L_{\mat} + d \ell$, and $\theta = \theta^g + \theta^\phi$, and in the second line, we have used the Cartan magic formula. Now, from covariance it follows  that
\begin{equation}
    \lie_{V_\zeta} \theta = \lie_{\zeta} \theta = d(i_\zeta \theta) + i_{\zeta} (d\theta).
\end{equation}
Using equation \eqref{eq:delL}, $\dl J_{\zeta}$ can be rewritten as
\begin{align}
    \begin{split} \label{eq:delJ}
        \dl J_{\zeta} &= d(i_\zeta \theta) + i_{\zeta} (d\theta) - I_{V_\zeta} \dl \theta - i_{\zeta} (E_{ab} \dl g^{ab} + E_{A} \dl \phi^A + d\theta) \\
        &= d \left( i_\zeta \theta \right) - I_{V_\zeta} \dl \theta - i_\zeta (E_{ab} \dl g^{ab} + E_{A} \dl \phi^A ).
    \end{split}
\end{align}
Now, recall that in section \ref{sec:boundary-charges}, it was shown that $J_{\zeta} = dQ_{\zeta} + C_{\zeta}$ where $C_{\zeta}$ is proportional to the equations of motion. Plugging this into equation \eqref{eq:delJ} and rearranging, we get the Hollands-Wald identity:
\begin{equation} \label{eq:HW-os}
    -I_{V_\zeta} \dl \theta = d \chi_\zeta + G_{\zeta},
\end{equation}
where we have defined the $(D-2, 1)$ form $\chi_{\zeta}$ and the $(D-1, 1)$ form $G_{\zeta}$ as
\begin{align}
    \chi_{\zeta} &= \left(\dl Q_\zeta - i_{\zeta} \theta \right), \label{eq:chi-def} \\
    G_{\zeta} &=  \dl C_{\zeta} + i_{\zeta} \left(E_{ab} \dl g^{ab} + E_A \dl \phi^A \right). \label{eq:G-def} 
\end{align}
Note that pulling back to phase space $\mc{P}$ kills the $G_{\zeta}$ term due to the equations of motion. Therefore, when evaluated on phase space, equation \eqref{eq:HW-os} reduces to
\begin{equation} \label{eq:HW-P}
    -I_{V_\zeta} \omega = d\chi_{\zeta}.
\end{equation}
Integrating both sides of the above on a Cauchy surface $\Sigma$ we obtain
\begin{equation} \label{eq:HW-P-integrated}
    -I_{V_\zeta} \Omega_{\Sigma} = \int_{\partial \Sigma} \chi_{\zeta}.
\end{equation}
Note that if $\zeta$ is a small diffeomorphism that vanishes asymptotically, the right hand side of the above equation vanishes. This is simply the fact that small diffeomorphisms are gauge symmetries of gravity and thus do not have any associated symplectic structure. 

On the other hand, if $\zeta$ asymptotically approaches a Killing vector, the symplectic form is non-vanishing. For example if $\Sigma$ is a complete Cauchy surface in an asymptotically AdS spacetime, then Dirichlet boundary conditions on the metric (i.e., fixing the non-normalizable part of the metric\footnote{From the boundary CFT point of view, this has the interpretation that no sources are turned on in the Lorentzian section at asymptotic infinity.}) imply $\theta^g \big|_{\partial \Sigma} = 0$. Moreover, $\theta^\phi|_{\partial \Sigma} = 0$ due to similar asymptotic boundary conditions on the matter fields. Therefore, the second term in equation \eqref{eq:chi-def} drops out and we have
\begin{equation}
    \int_{\partial \Sigma} \chi_{\zeta} = \int_{\partial \Sigma} \dl Q_{\zeta},
\end{equation}
which implies
\begin{equation}\label{eq:SCHWI}
    -I_{V_{\zeta}}\Omega_{\Sigma} = \delta H_{\Sigma}(\zeta).
\end{equation}
This is the familiar statement from classical mechanics that the charge $H_{\Sigma}(\zeta)$ generates the action of $V_{\zeta}$ on phase space. In other words, the vector field $V_{\zeta}$ is a Hamiltonian vector field for the charge $H_{\Sigma}(\zeta)$.

Similarly, let $r$ be a homology surface in the entanglement wedge associated to the right boundary of an asymptotically AdS Schwarzchild black hole with the extremal surface $X$ such that $\partial r = X \cup R$. Note that in this case is the surface $r$ itself fluctuates as the metric fluctuates. This is why it is convenient to fix a gauge such that the coordinate location of the extremal surface is independent of the choice of metric. Such a gauge fixing is always possible, and is sometimes called the \emph{Hollands-Wald gauge} \cite{Hollands:2012sf, Lashkari:2015hha, Faulkner:2017tkh}. If $\xi$ is the asymptotic Killing vector field that vanishes at $X$, then $i_{\xi}\theta \big|_{X} = 0$ (since $\xi$ vanishes at $X$), and so
\begin{equation}
    \int_{\partial r} \chi_{\xi} = \int_{X} \dl Q_{\xi} - \int_{R} \dl Q_{\xi}.
\end{equation}
Therefore, in this gauge, the Hollands-Wald identity takes the form
\begin{equation}
    -I_{V_{\xi}} \Omega_r = \dl \left( \int_X Q_{\xi} - \int_R Q_{\xi} \right) = \dl H_{r}(\xi).
\end{equation}
Since all stationary black holes with bifurcate Killing horizons admit a global Killing vector field $\xi$, the above identity holds for all such black holes.

\section{Covariant phase space with quantum matter} \label{sec:cov-quant}
In this section we will generalize the results of section \ref{sec:prelim} to semi-classical gravity. This means that the matter degrees of freedom will be treated quantum mechanically, while gravity will be treated classically and taken to satisfy the semi-classical Einstein equation \cite{Flanagan:1996gw, Wall:2011hj}:
\begin{equation} \label{scEe}
    R_{ab} - \frac{1}{2} R g_{ab} + \Lambda g_{ab} = 8\pi G_N\, \langle \psi| T_{ab}| \psi\rangle,
\end{equation}
where $\psi$ is the state of bulk matter fields. Usually, it is inconsistent to treat a system part classically and part quantum mechanically, so to be clear, what is really being said here is that the quantum state of the metric can be well-approximated by a coherent state, while that of the matter fields can be more general. Coherent states are completely specified by the expectation value of a quantum observable, in this case, the metric. The semi-classical Einstein equation \eqref{scEe} together with matter dynamics specifies this expectation value. In the AdS/CFT context, this comes about as follows: we specify the boundary conditions at asymptotic infinity in terms of sources for the stress tensor and other single-trace or multi-trace operators. We will assume that there are no $O(N^2)$ sources for single/multi-trace operators, other than the stress tensor. Such sources would correspond to turning on classical background configurations for matter fields, which was already considered in section \ref{sec:prelim}. Our matter sources will be $O(1)$, which we will think of as turning on a quantum state $\psi^{(0)}$ for the bulk matter fields. Then, we solve the semi-classical Einstein equation perturbatively in $G_N$:
\begin{equation}\label{formal}
    g_{ab} = g^{(0)}_{ab} + G_N g^{(1)}_{ab} + \cdots, \psi = \psi^{(0)} + G_N \psi^{(1)}+ \cdots,
\end{equation}
where $g^{(0)}$ is the background metric obtained by solving Einstein's equations without matter but subject to the asymptotic boundary conditions for the metric determined by the source for the boundary stress tensor, $\psi^{(0)}$ is the quantum state of matter fields obtained by performing the matter path integral on $g^{(0)}$ subject to the matter boundary conditions (corresponding to $O(1)$ sources for single/multi-trace operators), $g^{(1)}$ is the $O(G_N)$ correction to the background metric from backreaction due to the stress tensor expectation value in $\psi^{(0)}$, $\psi^{(1)}$ is the change in the state of quantum fields owing to the backreaction on the metric, and so on. An important point that we have ignored in this analysis in the contribution of gravitons, or equivalently non-coherent deformations to the quantum state of the metric field. A somewhat hand-wavy justification for this is that we can treat gravitons as another quantum field, and take into account their back-reaction on the coherent part of the metric. A second, slightly more satisfactory way to deal with this issue is to consider a matter theory with a large central charge $c$, so that one can take $G_N \to 0$ in order to suppress quantum gravitational effects, while still retaining back-reaction, controlled by the parameter $\epsilon = c\,G_N$; in this case, the formal expansions in equation \eqref{formal} are to be treated as expansions in $\epsilon$, rather than in $G_N$ \cite{Hartle:1981zt}. We will not try to justify the semi-classical approximation any further than this; we will simply assume that in some appropriate limit, one can reliably use it. 

\subsection{Berry curvature} \label{sec:berry}
In the classical setting, the covariant phase space formalism naturally involves the symplectic form on phase space for both the metric and matter fields. However, in treating matter quantum mechanically, we will need to find a suitable generalization of the matter symplectic form. We propose that the matter Berry curvature is the appropriate generalization in this setting (see also \cite{Czech:2017zfq, Czech:2018kvg, Belin:2018bpg, Belin:2018fxe, Kirklin:2019ror, DeBoer:2019kdj, Parrikar:2023lfr, Czech:2023zmq, deBoer:2025rxx} for previous work involving the Berry curvature in gravity). 

To this end, let $\mc{P}_{\text{q}}$ be the space of solutions to the semi-classical Einstein equation. We assume that $\mathcal{P}_{\text{q}}$ is a manifold with coordinates $\{\lambda^i\}$, where each point corresponds to a solution $(g_{ab}(\lambda^i),\psi(\lambda^i))$. As before, let $\dl$ be the exterior derivative on $\mc{P}_{\text{q}}$ which can be expressed in terms of the coordinates $\{ \lambda^i \}$ as
\begin{equation} 
    \dl = \sum_i d\lambda^i \frac{\partial}{\partial \lambda^i}.
\end{equation}
The Berry connection is defined as the following $1$-form on $\mc{P}_{\text{q}}$:\footnote{Formally, consider the (trivial) bundle $\mc{G}=\mc{P}_{\text{q}} \times \mc{H}_{\text{matter}}$. Given a smooth section $|\psi\rangle$, we can smoothly associate a projector $P_{\psi}$ at each point on $\mc{P}_{\text{q}}$. Then $\ba$ is a $U(1)$ connection on the projected sub-bundle $P_{\Psi} \mc{G}$. } 
\begin{equation}\label{eq:Ber-conn}
    \ba = - i\langle \psi | \dl \psi\rangle.
\end{equation}
Under the action of the gauge group $|\psi\rangle \to e^{i\alpha} |\psi\rangle$, the connection transforms as
\begin{equation}
    \ba \to \ba' = \ba + \dl \alpha.
\end{equation}
The curvature associated to the connection $\ba$ is called the Berry curvature and is defined as
\begin{equation} \label{eq:Ber-curv}
    \bF = \dl \ba.
\end{equation}
The Berry curvature should be thought of as the natural generalization of the matter sympelctic 2-form over classical phase space. Indeed, if we take the family $|\psi(\lambda^i)\rangle$ to be a family of coherent states, then the Berry curvature reduces to the symplectic 2-form. To see this, consider a single scalar field $\widehat{\phi}$ with conjugate momentum $\widehat{\pi}$. Let $\Sigma$ be a Cauchy slice and let $(\phi(\lambda^i), \pi(\lambda^i))$ be a family of phase space configurations defined on $\Sigma$. Consider the family of coherent states $|\psi(\lambda^i)\rangle$ defined as:
\begin{equation} \label{eq:coh}
    |\psi(\lambda^i) \rangle = \exp \left[ i \int_{\Sigma} \epsilon_{\Sigma} \left( \pi(\lambda^i) \widehat{\phi} - \phi(\lambda^i) \widehat{\pi} \right)\right] |\Omega\rangle.
\end{equation}
These coherent states, by definition, satisfy:
\begin{equation}\label{eq:expec}
    \langle \psi |\, \widehat{\phi}\, |\psi \rangle = \phi,\quad  
    \langle \psi |\, \widehat{\pi}\, | \psi\rangle = \pi.
\end{equation}
From equation \eqref{eq:Ber-conn} it follows that the Berry connection takes the following form:
\begin{align}
    \begin{split}
        \ba &= -i \langle \psi | \dl  \psi \rangle \\
        &=  \int_{\Sigma} \eps_{\Sigma} \left( \dl \pi\, \phi - \dl \phi\, \pi \right),
    \end{split}
\end{align}
where in the second line we have used equation \eqref{eq:expec}. As a result, the Berry curvature for coherent states reduces to the classical symplectic form:
\begin{align}
    \begin{split}
        \bF = 2\int_{\Sigma} \epsilon_{\Sigma}\, \dl \pi \cw \dl \phi.
    \end{split}
\end{align}
This indicates that the natural generalization of the matter symplectic form to the setting of semi-classical gravity is the Berry curvature for matter states. We will show that with this choice, the covariant phase space formalism can be naturally generalized to semi-classical gravity.

\subsection{Semi-classical generalization of the symplectic form} 
To generalize the classical covariant phase space to the semi-classical setting, we assume that we have a family $\mc{P}_{\text{q}}$ of solutions $(g_{ab}(\lambda^i),\psi(\lambda^i))$ to the semi-classical Einstein equation:
\begin{equation}
    R_{ab}(g(\lambda)) - \frac{1}{2} g_{ab}(\lambda) R(g(\lambda)) + \Lambda g_{ab}(\lambda) = 8\pi G_N \langle \psi(\lambda) | T_{ab} | \psi(\lambda) \rangle.
\end{equation} 
\begin{figure}
    \centering
    \includegraphics[height=6.5cm]{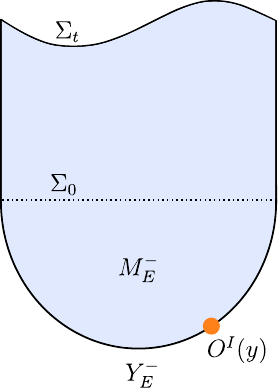}
    \caption{The bulk matter states are obtained by adding sources for single/multi-trace operators on the past asymptotic boundary $Y^-$.}
    \label{fig:EPI}
\end{figure}
We will assume that the matter states $\psi(\lambda^i)$ are prepared via a Euclidean path integral, potentially also including some Lorentzian time evolution. In the AdS/CFT context, this is a very natural assumption; in that context, the coordinates $\lambda^i$ can be thought of as some parameters in the boundary single/multi-trace sources, and $\psi(\lambda^i)$ are the bulk matter states obtained by performing the bulk matter path integral subject to these boundary conditions (plus $O(G_N)$ effects coming from back-reaction). We can schematically denote the bulk matter states as:
\begin{equation} \label{eq:EPIState}
   \langle \phi |\psi(\lambda^i)\rangle \sim \frac{1}{Z} \int^{\varphi|_{\Sigma}=\phi} [D \varphi]_{M^-}\, e^{ - S[\varphi, g_{ab}(\lambda^i)]-\int_{Y^-}J^I(\lambda^i)O_I},
\end{equation}
where $M_-$ is the portion of the bulk spacetime enclosed between the complete Cauchy slice $\Sigma$ and the lower-half portion of the asymptotic boundary $Y^-$ (see figure \ref{fig:EPI}), $S[\varphi]$ is the Euclidean/Lorentzian action for the bulk matter fields collectively denoted by $\varphi$, and $J_I(\lambda^i)$ is a family of sources for the boundary single/multi-trace operators $O_I$. These sources enter the bulk path integral in the form of operator insertions on the past asymptotic boundary $Y^-$. 

In the semi-classical context, it is natural to consider the object:
\begin{equation}
    \bbF = \Omega_{\text{grav.}} + \bF,
\end{equation}
where recall that $\Omega_{\text{grav.}}$ is the gravitational sympelctic 2-form, and $\bF$ is the Berry curvature for matter fields. An intuitive way to think about this is that in the semi-classical limit, the metric is taken to be in a coherent state with the expectation value of the metric determined by the semi-classical Einstein equation. Thus, the full quantum state in the bulk is roughly given by $|g_{ab},\psi\rangle = |g_{ab}\rangle\otimes  |\psi\rangle$. For coherent states, the Berry curvature reduces to the symplectic 2-form on phase space. Thus, one can think of $\bbF$ as being the ``total'' Berry curvature in the bulk. 

The 2-form $\bbF$ is independent of the Cauchy slice in Lorentz signature. To see this, first consider the simpler case of the matter Berry curvature on a fixed background metric (which is taken to be independent of the $\lambda^i$ for the moment). In this case, if we have two complete Cauchy slices $\Sigma$ and $\Sigma'$ (ending at the same boundary slice in the asymptotically AdS context), then the corresponding matter states are related to each other by unitary evolution:
\begin{equation}
\psi_{\Sigma'}(\lambda^i) = U(\Sigma,\Sigma')\,\psi_{\Sigma}(\lambda^i) ,
\end{equation}
\begin{equation}
    U(\Sigma,\Sigma') = \mathcal{P}\exp\left[-i\int_0^t ds \,h(s)\right],\;\; h(s) = \int_{\Sigma_s}n^{a}\xi^b T_{ab}
\end{equation}
where $\Sigma_s$ is some foliation of the Lorentzian spacetime region between $\Sigma$ and $\Sigma'$. The key point is that the unitary $U$ above is independent of the parameters $\lambda^i$, and so drops out of the Berry connection and curvature. Now, let us consider the more general case where the metric $g_{ab}$ can (and does) depend on the parameters $\lambda^i$. In this case, the Berry connection picks up an extra contribution:
\begin{equation}
    \ba_{\Sigma'} = \ba_{\Sigma} +\frac{1}{2} \int_{M(\Sigma,\Sigma')}\epsilon_{M}\,\delta g_{ab}\langle \psi|T^{ab} | \psi\rangle,
\end{equation}
where $\delta g_{ab}= \sum_i d\lambda^i (\pa_{\lambda^i}g_{ab})$, and $M(\Sigma,\Sigma')$ is the region bounded by the two slices $\Sigma$ and $\Sigma'$. Using the semi-classical Einstein equation, the corresponding Berry curvatures differ by:
\begin{equation}
    \bF_{\Sigma'} = \bF_{\Sigma} +  \int_{M(\Sigma,\Sigma')}\delta g_{ab}\curlywedge \delta G^{ab}
\end{equation}
where $G_{ab}$ is proportional to the Einstein equation without matter. This extra term precisely cancels against $\Omega_{\text{grav.}}(\Sigma') - \Omega_{\text{grav.}}(\Sigma)$. 
\subsection{Boundary charges }\label{sec:charge-quant}

Now consider a one-parameter family of diffeomorphisms generated by a vector field $\zeta$ in $M$. For the class of Euclidean path integral states (potentially also involving Lorentzian time evolution) discussed above, a diffeomorphism acts as a deformation of the metric:
$$\delta g_{ab} = \lie_{\zeta}g_{ab},$$
and for our purposes it will suffice to only consider diffeomorphisms which vanish on the past Euclidean boundary where sources for matter operators are turned on.\footnote{Note however that the diffeomorphism can act non-trivially on the Lorentzian boundaries.} Such a diffeomorphism $\zeta$ can be represented as a vector field on phase space which acts as follows: 
\begin{equation} 
I_{V_{\zeta}}\delta g_{ab} =\lie_{\zeta}g_{ab},
\end{equation}
\begin{equation}\label{eq:IVdelPsi}
I_{V_\zeta} \delta |\psi\rangle = i \int_{\Sigma} \epsilon^a \zeta^b T_{ab} | \psi \rangle .
\end{equation}
From equation \eqref{eq:IVdelPsi} it follows that the contraction of the matter Berry connection with $V_{\zeta}$ is given by
\begin{equation} \label{eq:JmatQ}
    I_{V_{\zeta}} \ba = \int_{\Sigma} \epsilon^a \langle \psi | T_{ab} | \psi \rangle \zeta^b.
\end{equation}
This lets us define a semi-classical charge associated to the diffeomorphism $\zeta$ as
\begin{equation} \label{eq:QWZCharge}
    H_\Sigma(\zeta) = \int_{\Sigma} \left( I_{V_\zeta} \theta^g - i_\zeta L_{g} \right) + I_{V_\zeta} \ba,
\end{equation}
where $\theta^g$ is defined in equation \eqref{eq:LvarTerms} and $L_{g} = L_{\text{EH}} + d\ell$ is the gravitational Lagrangian including the Gibbons-Hawking term (plus possible counterterms for holographic renormalization). This should be thought of as a semi-classical generalization of the classical charge given in equation \eqref{eq:WZCurrent}. Note that although equation \eqref{eq:QWZCharge} naively looks non-local, it can be written as the integral of a local current $J_{\zeta}$ on $\Sigma$ as
\begin{align}
    J_{\zeta} = \left( I_{V_\zeta} \theta^g - i_\zeta L_{g} \right) + \epsilon_a \langle T^{ab} \rangle_{\psi} \zeta_b.
\end{align}

To show that the charge $H_\Sigma(\zeta)$ is conserved (on-shell) for small diffeomorphisms that vanish asymptotically, we need to show that the local current $J_{\zeta}$ is conserved. From the definition of the current, we have
\begin{equation} \label{eq:Qdiff}
    dJ_{\zeta} = d(I_{V_\zeta} \theta^g - i_\zeta L_{g}) + \epsilon_M \nabla_a\left( \langle T^{ab} \rangle_{\psi} \zeta_b \right).
\end{equation}
Using diffeomorphism covariance of the classical gravitational Lagrangian, one can show that the gravitational terms above can be written as (see equation \eqref{eq:dJ})
\begin{equation} \label{eq:QChargeG}
     d(I_{V_\zeta} \theta^g - i_\zeta L_{g}) =  -\eps_M (2 G_{ab} \nabla^a \zeta^b).
\end{equation}
where $G_{ab}$ is the Einstein tensor with cosmological constant. Using the conservation of the matter stress tensor, the matter terms in $dJ$ can be written  as
\begin{equation}
    \epsilon_M \nabla^a \left( \langle T_{ab}\rangle_{\psi}  \zeta^b \right)=  \epsilon_M \langle T_{ab} \rangle_{\psi} \nabla^a \zeta^b.
\end{equation}
Combining this with equation \eqref{eq:QChargeG} we find that
\begin{equation}
    d J_{\zeta} = - 2  E_{ab}^{\text{SC}} \nabla^a \zeta^b,
\end{equation}
where $E^{\text{SC}}_{ab}$ is the full semi-classical equation of motion:
\begin{equation} \label{eq:ESC}
    E^{\text{SC}}_{ab} = \frac{1}{16\pi G_N} \epsilon_M \left( R_{ab} - \frac{1}{2} g_{ab} R + \Lambda g_{ab} - 8\pi G_N \langle \psi | T_{ab} | \psi \rangle \right).
\end{equation}
Therefore $H_{\Sigma}(\zeta)$ is conserved on-shell.

Just as in the classical case, the charge can be explicitly written as (see Appendix \ref{app:deriv})
\begin{equation} \label{eq:WZQ}
    H_\Sigma(\zeta) = \int_{\partial \Sigma} Q_{\zeta} + \int_{\Sigma} C_{\zeta}^{\text{SC}},
\end{equation}
where $Q_{\zeta}$ is given by eq. \eqref{eq:Charge-EH} and $C_\zeta^{\text{SC}}$ is defined as
\begin{equation} \label{eq:CSC}
    C_{\zeta}^{SC} = \frac{1}{16\pi G_N} \epsilon^a \left( R_{ab} - \frac{1}{2} g_{ab} R + \Lambda g_{ab} - 8\pi G_N \langle \psi | T_{ab} | \psi \rangle \right) \zeta^b.
\end{equation}
Therefore, as long as $\zeta$ vanishes sufficiently fast near $\partial \Sigma$ and the semi-classical Equations of motion hold, $H_\Sigma(\zeta)$ triviall vanishes. On the other hand, if $\zeta$ is a ``large'' diffeomorphism, then $H_\Sigma(\zeta)$ gives a non-trivial charge which is conserved provided $\zeta$ asymptotically approaches a Killing vector field; the argument for this follows along similar lines as the classical version explained previously.

\subsection{The semi-classical Hollands-Wald identity } \label{sec:quant-HW}
We will now generalize the classical Hollands-Wald identity, equation \eqref{eq:SCHWI}, by including quantum matter. The first step is to generalize the notion of diffeomorphism covariance to the semi-classical setting. To motivate our construction, we will first restrict to the class of Euclidean path integral states with sources inserted at the asymptotic boundary. For these states, the matter Berry connection $\ba$ is given by the following path integral: 
\begin{equation}
     \ba = \int [D\varphi]_{M^-\cup M^+} \int_{M^-} \epsilon\, \left( \frac{1}{2} \dl g^{ab} T_{ab} +\cdots \right) e^{ - S[\varphi,g] - \int_{Y}  J^I O_I},
\end{equation}
where the path integral is performed over the full contour (with a potential time-fold due to Lorentzian time evolution), the stress tensor insertion is only in the lower half portion corresponding to the ket, and the dots denotes other matter contributions which localized to the past asymptotic boundary. Consider now the object $\lie_{V_\zeta}\ba$. We will assume the following generalization of the classical notion of diffeomorphism covariance to the semi-classical setting: the phase space Lie derivative $\lie_{V_\zeta}$ can be replaced with the spacetime Lie derivative $\lie_{\zeta}$ inside the path integral. This will of course be the case if the matter action and the path integral measure are diffeomorphism covariant. Thus,
\begin{equation}
     \lie_{V_\zeta}\ba = \int [D\varphi]_{M^-\cup M^+} \int_{M^-} \lie_{\zeta}\left( \frac{1}{2}\epsilon \,  T_{ab} \dl g^{ab}  \right) e^{ - S[\varphi,g] - \int_{Y}  J^I O_I},
\end{equation}
where we have dropped the matter terms since we will focus on diffeomorphisms which act trivially near the past asymptotic boundary. Using Cartan's magic formula on $M^-$, the Lie derivative acting on a top form gives a total derivative. Thus, we obtain
\begin{equation}\label{eq:Qcov} 
    \lie_{V_{\zeta}} \ba = \frac{1}{2} \int_{\Sigma} i_{\zeta} \epsilon\, \langle \psi | T_{ab} | \psi \rangle \dl g^{ab} . 
\end{equation}
In the following we will take equation \eqref{eq:Qcov} as the \emph{definition} of the Lie derivative of $\ba$. In other words, we demand that equation \eqref{eq:Qcov} continues to hold for non-Euclidean path integral states. Note also that using Cartan's magic formula on phase space, the above Lie derivative can be written as 
\begin{equation} \label{eq:QCartan}
    \lie_{V_{\zeta}} \ba = I_{V_{\zeta}} \bF + \dl ( I_{V_{\zeta}} \ba).
\end{equation}

We are now in a position to generalize the classical Hollands-Wald idenitity to the semi-classical setting. Taking a phase space exterior derivative of the semi-classical charge defined in equation \eqref{eq:QWZCharge}, we obtain
\begin{equation} \label{eq:delQSCGlobal}
    \delta H_{\Sigma}(\zeta) = \int_{\Sigma} \delta (I_{V_\zeta} \theta^g - i_\zeta L_{g}) + \dl (I_{V_\zeta} \ba).
\end{equation}
Repeating the analysis of section \ref{sec:class-HW} we find that the gravitational terms yield
\begin{equation} \label{eq:dlQSCgrav}
    \int_{\Sigma} \dl (I_{V_\zeta} \theta^g - i_\zeta L_{g}) =\int_{\partial \Sigma} i_\zeta \theta^g - \int_{\Sigma} I_{V_\zeta} \dl \theta^g - \int_{\Sigma} i_\zeta G_{ab} \delta g^{ab},
\end{equation}
where $G_{ab}$ is the Einstein tensor (including the cosmological constant term). Recall that the above computation crucially uses diffeomorphism covariance of the classical gravitational action. We now process the matter term in an analogous fashion by using our semi-classical generalization of diffeomorphism covariance. Indeed, using equations \eqref{eq:Qcov} and \eqref{eq:QCartan}, we get
\begin{eqnarray} \label{eq:dlQSCmatter}
\dl (I_{V_\zeta} \ba) &=&  \lie_{V_{\zeta}} \ba - I_{V_{\zeta}}\bF\nonumber\\
&=& \frac{1}{2} \int_{\Sigma} i_{\zeta} \epsilon\, \langle \psi | T_{ab} | \psi \rangle \dl g^{ab} -I_{V_{\zeta}}\bF  .
\end{eqnarray}
Substituting equations \eqref{eq:dlQSCgrav} and \eqref{eq:dlQSCmatter} in equation \eqref{eq:delQSCGlobal} and  rearranging, we get:
\begin{equation}
 - I_{V_\zeta} \bbF = \int_{\partial \Sigma} \chi_{\zeta} + \int_{\Sigma} G_{\zeta}^{\text{SC}},
\end{equation}
where recall that we have defined
\begin{equation}
   \bbF = \Omega_{\text{grav.}} + \bF,
\end{equation}
\begin{equation} \label{eq:GSC}
    G_{\zeta}^{\text{SC}} = \int_{\Sigma} \left( \dl C_{\zeta}^{\text{SC}} + i_\zeta E_{ab}^{\text{SC}} \delta g^{ab} \right),
\end{equation}
with $C^{\text{SC}}_{\zeta}$ and $E^{\text{SC}}_{ab}$ being defined in equations \eqref{eq:CSC} and \eqref{eq:ESC} respectively. Just as in the classical case, as long as the semi-classical equation of motion holds, the quantity on the right hand side is a pure boundary term. Using the same arguments we gave in the classical case, this boundary term is given by $\delta H_{\Sigma}$, and so we get:
\begin{equation} 
- I_{V_\zeta} \bbF = \delta H_{\Sigma}.
\end{equation}
We can think of this as the semi-classical generalization of the Hollands-Wald identity. In particular, this also implies
\begin{equation}
    \lie_{V_{\zeta}}\bbF = 0,
\end{equation}
so that the semi-classical symplectic form is preserved under the action of diffeomorphisms on phase space.

\section{Covariant phase space formalism for subregions} \label{sec:quant-HW-sub}
In this section we will generalize our considerations of the covariant phase space formalism in the presence of quantum matter to the case of subregions of a Cauchy slice. Our discussion in this section will be limited to the case of small perturbations around a background black hole solution. 

\subsection{Purifications and the subregion Berry connection in QFT} \label{sec:purifications}
In this section, we will discuss how to define the Berry connection for subregions in quantum field theory on a \emph{fixed} background spacetime; we will return to the case of gravity in the next subsection. Subregions in QFT are associated not with pure states, but with mixed states, and there does not appear to be a standard way to associate a Berry connection with a family of mixed states (see \cite{Kirklin:2019ror, Czech:2017zfq, Czech:2018kvg, DeBoer:2019kdj, Parrikar:2023lfr, Czech:2023zmq, deBoer:2025rxx} for previous work on this). Here, we will consider a Berry connection for mixed states on subregions by constructing a specific purification involving the Connes cocycle, and defining the Berry connection using the corresponding purified global states. 

Consider a spacetime $M$ with a fixed background metric $g^{(0)}_{ab}$. Let $D(r)$ and $D(\bar{r})$ be two complementary subregions, defined in a gauge invariant way. For instance, we could take $D(r)$ to be the entanglement wedge of the boundary region $R$. Let $\Sigma$ be a Cauchy slice that passes through the quantum extremal surface corresponding to $R$, and let $r \subseteq \Sigma$ be the part of $\Sigma$ in $D(r)$. Let $\mc{M}_r$ be the von Neumann algebra of observables associated to the domain of dependence $D(r)$. Likewise, let $\overline{r} = \Sigma \setminus r$. Let $\Omega$ be a cyclic and separating state with respect to $\mc{M}_r$. For instance, if $M$ is an AdS black hole, we can take $\Omega$ to be the Hartle-Hawking state. We wish to define a Berry connection for density matrices on $r$. For simplicity, we can imagine that we have chosen some UV regulator such that the global Hilbert space factorizes into the Hilbert space of $r$ and Hilbert space of $\overline{r}$:
$$\mathcal{H} = \mathcal{H}_r\otimes \mathcal{H}_{\bar{r}},$$
but much of our discussion can be stated entirely in algebraic terms. Let $\psi$ be some global excited state and let $\rho_{\psi}$ be the corresponding density matrix on $r$. Our goal is to construct a purification of $\rho_{\psi}$ which we can use to define a subregion notion of the Berry connection; for this, the purification should be such that on $D(\overline{r}),$ the corresponding state looks like the Hartle-Hawking vacuum. Roughly speaking, this purification is constructed by ``sewing'' the local density matrix $\rho_{\psi}$ with the vacuum density matrix $\rho'_{\Omega}$ copy of the  on the complementary region. Note that in finite dimensional quantum systems, two local density matrices cannot be sewn into a global pure state unless they have the same Schmidt spectrum. In contrast, quantum field theory has the remarkable property that any pair of local states can be sewn into a global pure state via the action of local unitaries. This result, called the \emph{Connes-St\o rmer theorem}, arises from the fact that local algebras associated to subregions in quantum field theory are type III$_1$ \cite{connes1978homogeneity}. A pedagogical account of this result can be found in \cite{Lashkari:2019ixo}.

While the Connes-St\o rmer construction guarantees the existence of a purification that sews two local states together, it does not give a practical way of constructing the local unitaries which do the job. In states which look approximately stationary at late times, there is an alternate, somewhat more explicit construction which uses the \emph{Connes cocyle}: 
\begin{equation}
\widehat{\psi}_s = u'_{\Omega| \psi}(s)\, \psi,
\end{equation}
where
\begin{equation}
    u'_{\Omega | \psi}(s) = \Delta_{\Omega|\psi}^{\prime\, is} \Delta_{\psi}^{\prime\, -is}
\end{equation}
is the Connes-cocyle. Here, $\Delta'_{\psi} = \rho^{-1}_{\psi}\otimes \rho'_{\psi}$ where $\rho_{\psi}$ is the density matrix of $\psi$ on $r$, while $\rho'_{\psi}$ is the density matrix on $\overline{r}$. Similarly, $\Delta'_{\Omega|\psi} = \rho^{-1}_{\psi}\otimes \rho'_{\Omega}$ is the relative modular operator, and so 
\begin{equation}\label{cocyclefact}
    u'_{\Omega|\psi}(s) = \mathbb{1} \otimes (\rho'_{\Omega})^{is}(\rho'_{\psi})^{-is}.
\end{equation}

Using standard properties of modular flow, it is easy to check that
\begin{equation}
    \langle \widehat{\psi}_s| a |\widehat{\psi}_s\rangle = \langle \psi|a|\psi\rangle,\qquad \langle \widehat{\psi}_s| a'|\widehat{\psi}_s\rangle = \langle \psi|\Delta_{\Omega}^{-is}a'\Delta_{\Omega}^{is}|\psi\rangle.
\end{equation}
If $\psi$ is stationary at late times, i.e., satisfies
\begin{equation}
    \lim_{s\to \infty}\langle \psi|\Delta_{\Omega}^{-is}a'\Delta_{\Omega}^{is}|\psi\rangle = \omega'(a),
\end{equation}
then $\widehat{\psi}_s$ in the limit $s\to \infty$ approaches a purification of $\psi$ which corresponds to $\Omega$ on the complementary region. In this way, the Connes-cocycle gives a practical way to construct the desired purifications of local states. In \cite{Lashkari:2019ixo}, it was indeed argued that the infinite time limit of the Connes cocycle flow gives an approximate implementation of the Connes-St\o rmer construction with respect to a weaker distance measure (see also \cite{Parrikar:2024zbb} for some further discussion).  In the rest of this paper, we will focus on these more practical purifications involving the Connes-cocycle. We will use this purification to define a notion of the subregion Berry connection:
\begin{equation}\label{subBerry}
\ba_r = \lim_{s\to \infty} -i\langle \widehat{\psi}_s | \delta \widehat{\psi}_s\rangle.
\end{equation}

The Berry connection defined this way is independent of deformations of the Cauchy slice localized within the domain of dependence $D(r)$ and $D(\overline{r})$. Indeed, a deformation of the Cauchy surface localized within the region $D(r)$ and $D(\overline{r})$ can be thought of as a factorized local unitary $\mathcal{U} = U_r\otimes V_{\bar{r}}$ acting on the global state $\psi$, and similarly on $\Omega$. It is a straightforward exercise to check using equation \eqref{cocyclefact} that under the action of this unitary,
\begin{equation}
 \widehat{\psi}_s \to \mathcal{U}\, \widehat{\psi}_s,
\end{equation}
and since $\delta \mathcal{U}=0$, the Berry connection is independent of the choice of Cauchy slice.

\subsection{The subregion Berry connection in semi-classical gravity}
Now consider a family of solutions $(g_{ab}(\lambda_i), \psi(\lambda^i))$ to the semi-classical Einstein equation, where $\psi$ are global states; for concreteness, let us consider spacetimes with the same topology as the two-sided black hole. Consider the entanglement wedge $D(r)$ corresponding to the boundary region $R$, and let $X$ be the quantum extremal surface. Let $\Sigma$ be a Cauchy surface that passes through the extremal surface and let $r$ be a homology surface between $X$ and the asymptotic boundary region $R$. Likewise, let $\overline{r} = \Sigma \setminus r$ be the homology surface between the surface and the other asymptotic boundary. Our goal in this section is to associate a semi-classical phase space to the subregion $D(r)$ in the presence of quantum matter. One subtlety here is that the subregion $r$ itself depends on the parameters $\lambda^i$, since the location of the quantum extremal surface depends on the metric and the quantum state of matter fields. As was pointed out in section \ref{sec:class-HW}, the way to get around this is to fix the Hollands-Wald gauge, so that the coordinate location of the extremal surface $X$ is independent of the parameters $\lambda^i$.
With this choice, for the gravitational part, we can construct the subregion symplectic form by simply integrating the symplectic current $\omega_{\text{grav.}}$ over the subregion $r$. For the matter part, we propose that we take the Berry curvature of the subregion Berry connection defined in equation \eqref{subBerry}. Thus, our proposal for the semi-classical sympelctic form for the subregion $r$ is: 

\begin{equation}
    \bbF_r = \Omega_{\text{grav.},r} + \bF_r,
\end{equation}
where $\bF_r = \delta \ba_r$. In principle, the Hartle-Hawking vacuum state lives on the original black hole spacetime $g^{(0)}_{ab}$, while the excited state $\psi$ lives on the deformed metric $g_{ab}(\lambda^i)$. When $g_{ab}(\lambda^i)$ is a small deformation around $g^{(0)}_{ab}$, we expect that this distinction is not important, as the change in the background metric can be interpreted as a perturbation in the state around the HH vacuum. This highlights the inherently perturbative nature of our construction. Indeed, our construction involves a special background state, namely the HH state -- thus, it is not surprising that it only works perturbatively around this state and around the black hole geometry. It would be interesting to generalize this beyond the regime of small perturbations; we leave this for future work.

As before, $\bbF_r$ is invariant under deformations of the Cauchy slice localized within $D(r)$. This follows straightforwardly from the fact that such localized deformations act like localized unitary transformations within $D(r)$, together with the arguments similar to those given in the global case.

\subsection{A subregion Hollands-Wald identity}
We are now in a position to derive the subregion Hollands-Wald identity. Given a diffeomorphism $\zeta$, we define the subregion diffeomorphism charge as
\begin{equation} \label{eq:HSub}
    H_{r}(\zeta) = \int_r \left( I_{V_\zeta} \theta^g - i_{\zeta} L_g \right) + I_{V_\zeta} \ba_r.
\end{equation}
As before, the charge can be written as an integral of a current density over $r$. To see this, recall from equation \eqref{eq:IVdelPsi} that the action of $V_\zeta$ on the state $\hpsi$ is given by
\begin{equation} \label{eq:IVhpsi}
    I_{V_\zeta} | \dl \hpsi \rangle = i \int_{\Sigma} \epsilon^a \zeta^b T_{ab} | \hpsi \rangle
\end{equation}
where $\Sigma = r \cup \rbar$. Then, the matter term in equation \eqref{eq:HSub} becomes
\begin{align} \label{eq:IVar}
    \begin{split}
        I_{V_{\zeta}} \ba_r &= \int_{\Sigma} \epsilon^a \zeta^b \langle \hpsi | T_{ab} |\widehat{\psi} \rangle \\
        &= \int_{\rbar} \epsilon^a \zeta^b \omega'(T_{ab}) + \int_r \epsilon^a \zeta^b \psi(T_{ab}) \\
        &= \int_r \epsilon^a \zeta^b \langle T_{ab} \rangle_{\psi}
    \end{split}
\end{align}
where in the second line we have used the fact that the infinite time limit of the cocycle sews $\psi$ with $\omega$ in $\mc{M}'$, and in the third line we have used $\omega'(T_{ab}) = 0$. Therefore, the charge $H_{r}(\zeta)$ is the integral of the following current density over $r$:
\begin{equation}
    J_{\zeta} = I_{V_{\zeta}} \theta^g - i_{\zeta} L_g + \epsilon^a \langle T_{ab} \rangle_{\psi} \zeta^b.
\end{equation}
Just as in the global case, it can be shown that $d J_{\zeta} = 0$ as long as the semi-classical equations of motion hold. Therefore $H_{\Sigma}(\zeta)$ is conserved for small diffeomorphisms. However for large diffeomorphisms it reduces to a boundary term up to terms proportional to the equations of motion:
\begin{equation} \label{eq:dlHR}
    H_{R}(\zeta) = \int_{\partial r} Q_{\zeta} + \int_r C_{\zeta}^{\text{SC}},
\end{equation}
where $Q_{\zeta}$ and $C_{\zeta}^{\text{SC}}$ are defined in equations \eqref{eq:Charge-EH} and \eqref{eq:CSC}.

To derive the subregion Hollands-Wald identity, we take a variation of the charge:
\begin{align}
    \begin{split}
        \dl H_{r}(\zeta) &= \int_r \dl \left( I_{V_\zeta} \theta^g - i_\zeta L_g \right) + \dl ( I_{V_\zeta} \ba_r ) \\
        &= \int_{\partial r} i_{\zeta} \theta^g - \int_{r} I_{V_\zeta} \omega^g - \int_{r} i_{\zeta} G_{ab} \dl g^{ab} + \dl ( I_{V_\zeta} \ba_r).
    \end{split}\label{eq:CartanR} 
\end{align}
To process the matter term we use Cartan's magic formula:
\begin{equation}
    \dl (I_{V_\zeta} \ba_r) = \lie_{V_\zeta} \ba_r - I_{V_\zeta} \bF_r
\end{equation}
where recall that $\bF_r = \dl \ba_r$. Just like the charge, one can show that the Lie derivative of the subregion Berry connection is simply the restriction of the Lie derivative of the global Berry connection to the subregion. Indeed, from equation \eqref{eq:Qcov} we have
\begin{align} \label{eq:LieAR2}
    \begin{split}
        \lie_{V_\zeta} \ba_r &= \frac{1}{2} \int_{\Sigma} \zeta^c \epsilon_c\, \dl g^{ab} \langle \hpsi | T_{ab} | \hpsi \rangle \\
        &= \frac{1}{2} \int_r \zeta^c \epsilon_c\, \dl g^{ab} \psi(T_{ab}) + \frac{1}{2} \int_{\rbar} \zeta^c \epsilon_c\, \dl g^{ab} \omega'(T_{ab}) \\
        &= \frac{1}{2} \int_r \zeta^c \epsilon_c\, \dl g^{ab}\, \langle T_{ab}\rangle_{\psi}
    \end{split}
\end{align}
where again we have used the condition $\langle \hpsi |  T_{ab}(x) | \hpsi \rangle = \omega'(T_{ab}(x)) = 0$ whenever $x \in \rbar$. Combining equations \eqref{eq:dlHR}, \eqref{eq:CartanR}, and \eqref{eq:LieAR2} we obtain the subregion Hollands-Wald identity:
\begin{equation}
    -I_{V_{\zeta}} \bbF_r = \int_{\partial r} \chi_{\zeta} + \int_r G_{\zeta}^{\text{SC}},
\end{equation}
which reduces to 
\begin{equation} \label{eq:HWSC}
    -I_{V_{\zeta}} \bbF_r = \delta H_r(\zeta),
\end{equation}
upon using equations of motion. We emphasize that this result is only valid in the Hollands-Wald gauge. In going to a different gauge, one often picks up additional boundary terms localized at the quantum extremal surface \cite{Lashkari:2015hha}.

\subsection{Consistency with the generalized entropy prescription} \label{sec:HWPert}
The semi-classical subregion Hollands-Iyer-Wald identity, equation \eqref{eq:HWSC}, is valid only for small perturbations around a background metric. In this section, we will perturbatively expand the identity and explore some of its consequences. As before, we consider a family of solutions $(g_{ab}(\lambda), \psi(\lambda))$ to the semi-classical Einstein equation and expand around some chosen background $(g^\0_{ab}, \psi^\0)$. The expansion is performed simultaneously in $G_N$ and $\lambda$; consequently, each term is labeled by two indices, one indicating the power of $G_N$ and the other the power of $\lambda$. We thus expand the solution as
\begin{align}
    \psi &= \psi^\0 + \psi^{(1, 0)} + \psi^{(2, 1)} + \cdots \label{eq:exp1} \\
    g_{ab} &= g_{ab}^\0 + g_{ab}^{(1, 1)} + g_{ab}^{(2, 2)} + \cdots \label{eq:exp2}
\end{align}
where, in each term, the first index denotes the order in $\lambda$ and the second the order in $G_N$. Thus $\psi^{(1, 0)}$ is the $O(\lambda)$ change in the state, and $g^{(1,1)}$ is the $O(\lambda G_N)$ back-reaction sourced by $\psi^{(1,0)}$ via the linearized Einstein equation. Likewise, $\psi^{(2, 1)}$ is the $O( \lambda^2 G_N )$ change in the state induced by the metric perturbation, and $g^{(2, 2)}$ is the $O(\lambda^2 G_N^2)$ change in the metric sourced by $\psi^{(2, 1)}$ via the second order Einstein equation. The dots denote terms that are higher order in $\lambda$ and $G_N$.  

Let $\xi$ be the bulk boost isometry restricted to the two wedges $D(r) \cup D(\rbar)$. Once the semi-classical equations of motion are imposed, equation \eqref{eq:HWSC} takes the form
\begin{equation} \label{eq:HIWSub}
    \dl \langle K_{\text{CFT}} \rangle - \frac{\dl \mc{A}}{4G_N} = - I_{V_\xi} \Omega^{\text{grav}}_r - I_{V_\xi} \bF_r
\end{equation}
where $\dl \langle K_{\text{CFT}} \rangle$ is the change in the CFT modular Hamiltonian and $\dl \mc{A}$ is the change in the area of the extremal surface. Then, using equation \eqref{eq:IVhpsi}, the bulk matter term can be written as
\begin{align}
    \begin{split}
        -I_{V_\xi} \bF_{r} &= -i \left( \langle \dl \hpsi | I_{V_\xi} \dl \hpsi \rangle - \langle I_{V_\xi} \dl \hpsi | \dl \hpsi \rangle \right) \\
        &= \int_{\Sigma} \epsilon^a \xi^b \left( \langle \dl \hpsi | T_{ab} | \hpsi \rangle + \langle \hpsi | T_{ab} | \dl \hpsi \rangle \right) \\
    \end{split}
\end{align}
Now, it follows from definition that $|\dl \hpsi\rangle = \dl u' |\psi \rangle + u' | \dl \psi \rangle$ where $u'$ denotes the infinite time limit of CC flow in $\mc{M}'$. Then the matter Berry curvature becomes
\begin{align}
    \begin{split}
        -I_{V_\xi} \bF_r = \int_{\Sigma} \epsilon^a \zeta^b &\left( \langle \dl \psi | (u')^\dagger T_{ab} u' | \psi \rangle + \langle \psi | (u')^\dagger T_{ab} u' | \dl \psi \rangle \right. \\ & \left. + \langle \psi | (\dl u')^\dagger T_{ab} u' | \psi \rangle + \langle \psi | (u')^\dagger T_{ab} \dl u' | \psi \rangle \right).
    \end{split}
\end{align}
Next, we split up the integral over $\Sigma$ as a sum over $r$ and $\rbar$. It is easy to check that the terms localized on $\rbar$ take the form $\dl \int_{\rbar} \epsilon^a \xi^b \langle \hpsi | T_{ab} | \hpsi \rangle$. But $\langle \hpsi | T_{ab}(x) | \hpsi \rangle = \omega'(T_{ab}(x)) = 0$ for all $x \in \rbar$, and for all states $\psi$. These terms therefore vanish. To process the terms localized on $r$, note that $T_{ab}(x)$ commutes with $u'$ when $x \in r$. Then, using the fact that $\dl (u'^\dagger u') = 0$, we can rewrite the above as
\begin{equation}
    -I_{V_\xi} \bF_r = \int_{r} \epsilon^a \xi^b \left( \dl \langle T_{ab} \rangle_{\psi} + \langle \psi | [T_{ab}, (u')^\dagger \dl u' ] | \psi \rangle \right).
\end{equation}
Clearly, the second term vanishes and we are left with
\begin{equation} \label{eq:IVfT}
    -I_{V_\xi} \bF_r = \int_r \epsilon^a \xi^b \dl \langle T_{ab} \rangle_{\psi}.
\end{equation}
The above equation is a purely kinematical fact that is valid for all states $\psi$ perturbatively around a fixed background. Once we perturbatively gauge fix the subregion $r$, we may expand the right hand side above in powers of $\lambda$ and $G_N$ using the expansions given in equations \eqref{eq:exp1} and \eqref{eq:exp2}. In the following we will only consider state perturbations that are at most $O(\lambda)$ while including back-reaction up to $O(G_N^2)$, in which case it suffices to fix the location of the extremal surface up to $O(G_N)$ \cite{Hollands:2012sf}.

Let us now expand equation \eqref{eq:HIWSub} order by order. At first order (\emph{i.e.} at $O(\lambda)$) the location of the extremal surface does not change, and the change in the area is solely due to the change in the background metric. Moreover, the gravitational symplectic form term drops out since $\xi$ is a Killing vector of the background metric. Then equation \eqref{eq:HIWSub} becomes
\begin{equation}
    \langle K_{\text{CFT}} \rangle^\1 = \frac{\mc{A}^\1}{4 G_N} + \langle K_{\text{bulk}} \rangle^\1,
\end{equation}
where we have identified $\langle K_{\text{bulk}} \rangle^\1 = \int_r \epsilon^a \xi^b \langle T_{ab} \rangle^\1$ as the linearized change in the bulk modular Hamiltonian associated to the matter state $\psi$. This, along with the first law of entanglement implies that
\begin{equation}
    S^\1_{\text{CFT}} = \frac{\mc{A}^\1}{4 G_N} + S^\1_{\text{bulk}},
\end{equation}
which is precisely the FLM formula \cite{Faulkner:2013ana}. At second order (\emph{i.e.} at $O(\lambda^2 G_N)$) it is necessary to impose the Hollands-Wald gauge to ensure that the extremal surface does not fluctuate. Then, subtracting the $O(G_N)$ change in the bulk entanglement entropy from both sides of equation \eqref{eq:HIWSub} we get
\begin{equation} \label{eq:SrelEq2}
    S^\2_{\text{CFT}}(\Psi \| \Omega) = -(I_{V_\xi} \Omega^{\text{grav}}_r)^\2 + \left( \int_r \xi^a \left( \epsilon^b \langle T_{ab} \rangle \right)^\2 - S_{\text{bulk}}^\2 \right).
\end{equation}
where in the second line we have used the quantum extremal surface (QES) prescrption \cite{Engelhardt:2014gca} to identify $S_{\text{gen}}$ with the CFT entanglement entropy. Note that the terms in parentheses above differ from the matter relative entropy due to backreaction:
\begin{equation}
    \int_r \xi^a \left(\epsilon^b \langle T_{ab} \rangle \right)^\2 = \langle K_{\text{bulk}}\rangle^\2 + \int_r \xi^a (\epsilon^b)^\1 \langle T_{ab} \rangle^\1.
\end{equation}
For coherent (i.e. single trace) deformations of the CFT state, the entanglement entropy term $S^\2_{\text{bulk}}$ drops out\footnote{This is because any coherent state is unitarily equivalent to the vacuum state: $\rho_{r}^{\text{coherent}} = U_r \rho_r^\0 U_r^\dagger$, where $U_r =  e^{i \int_r (\pi \widehat{\phi} - \phi \widehat{\pi} )}$} and the Berry curvature reduces to the matter symplectic form. Therefore for coherent deformations of the CFT vacuum, eq.\ \eqref{eq:SrelEq2} reduces to
\begin{equation}
    S^\2_{\text{CFT}}(\Psi \| \Omega) = -I_{V_\xi}\Omega_r^\2,
\end{equation}
where $\Omega_r = \Omega_r^{\text{grav}} + \Omega_r^{\text{matter}}$ is the full bulk symplectic form \cite{Jafferis:2015del, Lashkari:2015hha}. However, for more general non-coherent (i.e.\ multi trace) deformations of the CFT vacuum, eq.\ \eqref{eq:SrelEq2} implies that the bulk and boundary relative entropies no longer match at $O(G_N)$.  Arguments in favor of the mismatch between the bulk and boundary relative entropies at subleading orders in $G_N$ were first given in \cite{Dong:2017xht} using the holographic replica trick. Here we have presented an alternate, manifestly Lorentzian derivation of the mismatch using covariant phase space techniques.

\section{Bulk vs.\ boundary Berry curvature in AdS/CFT}\label{sec:Ads/cft}
Consider a family of Euclidean path integral states $\Psi(\lambda^i)$ in a holographic CFT, where the $\lambda^i$ are now sources for various single/multi-trace operators in the CFT. It is natural to define the Berry connection and curvature in the boundary CFT as:
\begin{equation} 
\bbA_{\text{CFT}} = \langle \Psi | \delta \Psi\rangle, \;\;\bbF_{\text{CFT}}  = \delta\bbA_{\text{CFT}} .
\end{equation} 
Ignoring non-perturbative effects, these states correspond to bulk states $\psi(\lambda^i)$ for the bulk matter fields, together with a backreacted metric $g_{ab}(\lambda^i)$ which solves the semi-classical Einstein equation. One can think of the metric as being in a coherent state with the semi-classical solution determining the expectation value of the metric, and thus denote the bulk states collectively as $|g_{ab},\psi\rangle$. The AdS/CFT dictionary gives a map $V$ from the bulk semi-classical states $|g,\psi\rangle$ to the boundary states $|\Psi\rangle$. The map should be an isometry by the AdS/CFT dictionary, up to non-perturbative corrections. Of course, the map $V$ does not depend on the parameters $\lambda^i$. This immediately implies:
\begin{equation}
    \bbF_{\text{CFT}}  = \delta\langle \Psi | \delta \Psi\rangle = \delta \langle g,\psi | V^{\dagger}\delta\left( V|g,\psi\rangle\right) = \delta\langle g,\psi | \delta |g,\psi\rangle = \bbF_{\text{bulk}}.
\end{equation}
The equality of bulk and boundary Berry curvatures is an almost trivial corollary of the AdS/CFT dictionary. In the classical limit, the bulk Berry curvature reduces to the symplectic 2-form on the bulk classical phase space, and one finds that the bulk symplectic form is dual to the boundary Berry curvature \cite{Belin:2018fxe}. Beyond the classical limit, one must also account for quantum corrections in the bulk, and the bulk symplectic form naturally gets replaced -- within the semi-classical approximation -- by the sum of the gravitational symplectic form and the matter Berry curvature.   

\section{Discussion}
In this work, we have proposed a natural generalization of the covariant phase space formalism to the case of semi-classical gravity. We end with some open questions and future directions:
\begin{enumerate}
    \item While we have defined a semi-classical symplectic form on the space of solutions to the semi-classical Einstein equation, it would be nice to find some applications for this formalism. In a companion paper \cite{upcoming}, we will discuss one application of this formalism in understanding the Ryu-Takayanagi formula in the quantum regime \cite{Engelhardt:2018kcs} from Lorentzian methods.  
    \item Usually, the symplectic structure on phase space is a starting point for quantization. It is not clear to us that the semi-classical symplectic form defined here should play such a role in the quantization of the metric. It would be good to clarify this point further.
    \item Our discussion in the case of subregions was limited to small perturbations around a background solution. It would be good to generalize this beyond perturbation theory. Relatedly, it would be good to understand the boundary dual of the subregion semi-classical symplectic form in AdS/CFT (see \cite{Kirklin:2019ror, deBoer:2025rxx} for some work along this direction). 
\end{enumerate}

\section*{Acknowledgments}
We would like to thank Jan de Boer, Abhijit Gadde, Alok Laddha, Shiraz Minwalla, Ashoke Sen and Sandip Trivedi for helpful discussions and comments on an earlier version of this draft. We acknowledge support from the Department of
Atomic Energy, Government of India, under project identification number RTI 4002, and from the Infosys Endowment for the study of the Quantum Structure of Spacetime.

\appendix
\section{Boundary charges in Einstein gravity}  \label{app:deriv}
In this appendix we will derive equations \eqref{eq:charge-os} and \eqref{eq:Charge-EH}. We begin by separating the current defined in equation \eqref{eq:Jdef} into gravitational and matter terms:
\begin{equation}
    J_{\zeta} = J_{\zeta}^g + J_{\zeta}^\phi,
\end{equation}
where we have defined
\begin{align}
    J_{\zeta}^g &= I_{V_\zeta}\theta^g - i_\zeta (L_{\EH} + d\ell),  \label{eq:Jg}\\
    J_{\zeta}^\phi &= I_{V_\zeta} \theta^\phi - i_{\zeta} L_{\mat}, \label{eq:Jphi}
\end{align}
and $V_{\zeta}$ is the vector field defined in equation \eqref{eq:CSVF}. Let us now evaluate equations \eqref{eq:Jg} and \eqref{eq:Jphi} term by term. Recall from equation \eqref{eq:LvarTerms} that $\theta^g$ is a $(D-1, 1)$ form given by
\begin{equation}
    \theta^g = \frac{1}{16 \pi G_N} \left( \nabla_a \delta g^{ab} - \nabla^b \delta g\indices{^a_a} \right) \epsilon_b + \dl \ell.
\end{equation}
Taking the interior product with respect to $V_{\zeta}$ and using covariance of $\ell$, we get
\begin{align}
    \begin{split} \label{eq:A6}
        I_{V_\zeta} \theta^g &= \frac{1}{16\pi G_N} \left( \nabla_a (\lie_{\zeta} g)^{ab} - \nabla^b (\lie_\zeta g)\indices{^a_a} \right) \epsilon_b + \lie_{\zeta} \ell \\
        &= \frac{1}{16 \pi G_N} \left( \nabla_a \nabla^a \zeta^b + \nabla^a \nabla^b \zeta_a - 2 \nabla^b \nabla^a \zeta_a \right) \epsilon_b + \lie_{\zeta} \ell \\
        &= \frac{1}{8 \pi G_N} \left(  \nabla_a \nabla^{[a} \zeta^{b]} + \zeta_a R^{ab} \right) \epsilon_b + \lie_{\zeta} \ell,
    \end{split} 
\end{align}
where, to go from the second line to the third line, we have used the relation
\begin{equation}
    \nabla^b \nabla^a \zeta_a = \nabla^a \nabla^b \zeta_a - R^{ab}\zeta_a.
\end{equation}
Next, note that
\begin{align} 
\begin{split}
    i_{\zeta} (L_{\text{EH}} + d \ell) &= \frac{1}{16 \pi G_N} \zeta^a \epsilon_a \left(R - 2 \Lambda \right) + i_{\zeta} d \ell \\
    &= \frac{1}{8\pi G_N} \zeta_a \left(\frac{1}{2} g^{ab} R - \Lambda g^{ab} \right) \epsilon_b + i_{\zeta} d\ell. \label{eq:A8}
\end{split}
\end{align}
Therefore, combining equations \eqref{eq:A6} and \eqref{eq:A8} we find that 
\begin{equation} \label{eq:A9}
    J_{\zeta}^g = \frac{1}{8\pi G_N} \epsilon_b \nabla_a \nabla^{[a} \zeta^{b]} + d(i_\zeta \ell) + \frac{1}{8\pi G_N} \zeta_a \left( R^{ab} - \frac{1}{2} g^{ab} R + \Lambda g^{ab} \right) \epsilon_b.
\end{equation}
Note that the first two terms in the above equation are $d$-exact:
\begin{equation}
    \frac{1}{8 \pi G_N} \epsilon_b \nabla_a \nabla^{[a} \zeta^{b]} + d (i_\zeta \ell) = d \left( - \frac{1}{16\pi G_N} \star d\zeta + i_{\zeta} \ell \right),
\end{equation}
where, on the right hand side we have lowered the index of $\zeta$ to interpret it as a one-form.

Finally, note that the matter contribution to the current can be expressed in terms of the matter stress tensor as
\begin{equation} \label{eq:JmatCl}
    J^{\phi}_{\zeta} = - \zeta_a T^{ab} \epsilon_b.
\end{equation}
One way to see this is to note that the matter current $J^{\phi}_\zeta$ as defined in equation \eqref{eq:Jphi} is precisely what one would get from applying Noether's procedure to the matter Lagrangian $L_{\mat}$ in the background $g$. This can also be explicitly checked in examples. For instance, suppose the matter Lagrangian is of the form
\begin{equation}
    L_{\mat} = - \left( \frac{1}{2} g^{ab} \nabla_a \phi \nabla_b \phi + V(\phi) \right) \epsilon_M
\end{equation}
where $V(\phi)$ is some potential. Using $\theta^\phi = - \dl \phi \nabla^a \phi\, \epsilon_a$, we get
\begin{align}
    \begin{split} \label{eq:A13}
        J_{\zeta}^\phi &= I_{V_\zeta} \theta_{\phi} - i_{\zeta} L_{\mat} \\
        &= - \zeta_{a} \left( \nabla^a \phi \nabla^b \phi - g^{ab} \left( \frac{1}{2} g^{cd} \nabla_c \phi \nabla_d \phi + V(\phi) \right) \right) \epsilon_b \\
        &= - \zeta_a T^{ab}_{\text{matter}} \epsilon_b.
    \end{split}
\end{align}
Adding equation \eqref{eq:JmatCl} to equation \eqref{eq:A9} we find that the current takes the form
\begin{equation} \label{eq:A14}
    J^g_{\zeta} = dQ_{\zeta} + C_{\zeta}
\end{equation}
where
\begin{align}
    Q_{\zeta} &= - \frac{1}{16\pi G_N} \epsilon^{ab} \nabla_{a} \zeta_b + i_{\zeta} \ell \\
    C_{\zeta} &= \frac{1}{8\pi G_N} \zeta_a \left( R^{ab} - \frac{1}{2} g^{ab}R + \Lambda g^{ab} - 8\pi G_N T^{ab}_{\text{matter}} \right) \epsilon_b
\end{align}
Integrating both sides of equation \eqref{eq:A14}, we find equations \eqref{eq:charge-os} and \eqref{eq:Charge-EH}.

Next, let us show that the charge furnished by the integrating current defined above on a Cauchy slice is can be expressed in terms of the Brown-York stress tensor. Let $r = r^a \partial_a$ be the outward pointing normal to $\partial M$, and let $n = n^a \partial_a$ be the future directed timelike normal of $\partial \Sigma$ in $\partial M$. Let $\epsilon_{\partial \Sigma}$ be the diffeomorphism invariant volume form on $\partial \Sigma$. Then, from equation \eqref{eq:Charge-EH}, we have
\begin{equation}
    H_{\Sigma}(\zeta) = - \frac{1}{16 \pi G_N} \int_{\partial \Sigma} \left( r^a n^b - n^a r^b \right) \nabla_a \zeta_b\, \epsilon_{\partial \Sigma} + \int_{\partial \Sigma} i_{\zeta} \ell.
\end{equation}
Note that $\zeta$ is a Killing vector of $\partial M$, and therefore $\nabla_a \zeta_b = - \nabla_b \zeta_a$ when pulled back to $\partial M$. Moreover, $(h^{ab} r_a \zeta_b)|_{\partial M} = 0$. This lets us rewrite the above as
\begin{equation}
    H_{\Sigma}(\zeta) = -\frac{1}{8\pi G_N} \int_{\partial \Sigma} n^a \zeta^b K_{ab} \epsilon_{\partial \Sigma} + \int_{\partial \Sigma} i_{\zeta} \ell,
\end{equation}
where $K_{ab} = \nabla_a r_b$ is the extrinsic curvature of $\partial M$. Finally, recall that $\ell$ is the Gibbons-Hawking-York term defined in equation \eqref{eq:GHY}. Collecting everything together we recover equation \eqref{eq:BYcharge}:
\begin{equation}
    H_{\Sigma}(\zeta) = \frac{1}{8\pi G_N} \int_{\partial \Sigma} \epsilon_a \left( K^{ab} - h^{ab} K \right) \zeta_b.
\end{equation}
Note that there are additional counterterms coming from $\ell$ that we have suppressed. These counterterms are necessary to render the charge finite \cite{Balasubramanian:1999re}.

\bibliography{refs}

@article{Engelhardt:2014gca,
    author = "Engelhardt, Netta and Wall, Aron C.",
    title = "{Quantum Extremal Surfaces: Holographic Entanglement Entropy beyond the Classical Regime}",
    eprint = "1408.3203",
    archivePrefix = "arXiv",
    primaryClass = "hep-th",
    doi = "10.1007/JHEP01(2015)073",
    journal = "JHEP",
    volume = "01",
    pages = "073",
    year = "2015"
}

@article{Shi:2020csw,
    author = "Shi, Kai and Wang, Xuan and Xiu, Yihong and Zhang, Hongbao",
    title = "{Covariant phase space with null boundaries}",
    eprint = "2008.10551",
    archivePrefix = "arXiv",
    primaryClass = "hep-th",
    doi = "10.1088/1572-9494/ac2a1b",
    journal = "Commun. Theor. Phys.",
    volume = "73",
    number = "12",
    pages = "125401",
    year = "2021"
}

@article{Bousso:2025xyc,
    author = "Bousso, Raphael",
    title = "{Robust Singularity Theorem}",
    eprint = "2501.17910",
    archivePrefix = "arXiv",
    primaryClass = "hep-th",
    doi = "10.1103/6f9b-3jmx",
    journal = "Phys. Rev. Lett.",
    volume = "135",
    number = "1",
    pages = "011501",
    year = "2025"
}

@article{Penington:2019npb,
    author = "Penington, Geoffrey",
    title = "{Entanglement Wedge Reconstruction and the Information Paradox}",
    eprint = "1905.08255",
    archivePrefix = "arXiv",
    primaryClass = "hep-th",
    doi = "10.1007/JHEP09(2020)002",
    journal = "JHEP",
    volume = "09",
    pages = "002",
    year = "2020"
}

@article{Almheiri:2019psf,
    author = "Almheiri, Ahmed and Engelhardt, Netta and Marolf, Donald and Maxfield, Henry",
    title = "{The entropy of bulk quantum fields and the entanglement wedge of an evaporating black hole}",
    eprint = "1905.08762",
    archivePrefix = "arXiv",
    primaryClass = "hep-th",
    doi = "10.1007/JHEP12(2019)063",
    journal = "JHEP",
    volume = "12",
    pages = "063",
    year = "2019"
}

@article{Parrikar:2023lfr,
    author = "Parrikar, Onkar and Singh, Vivek",
    title = "{Canonical purification and the quantum extremal shock}",
    eprint = "2302.14318",
    archivePrefix = "arXiv",
    primaryClass = "hep-th",
    doi = "10.1007/JHEP08(2023)155",
    journal = "JHEP",
    volume = "08",
    pages = "155",
    year = "2023"
}

@article{DeBoer:2019kdj,
    author = "De Boer, Jan and Lamprou, Lampros",
    title = "{Holographic Order from Modular Chaos}",
    eprint = "1912.02810",
    archivePrefix = "arXiv",
    primaryClass = "hep-th",
    doi = "10.1007/JHEP06(2020)024",
    journal = "JHEP",
    volume = "06",
    pages = "024",
    year = "2020"
}

@article{upcoming,
    author = "Bhattacharya, Abhirup and Parrikar, Onkar",
    title = "{Towards a Lorentzian approach to quantum extremality}",
    journal = "to appear shortly"
}

@article{Kim:2023sig,
    author = "Kim, Seok and Kundu, Suman and Lee, Eunwoo and Lee, Jaeha and Minwalla, Shiraz and Patel, Chintan",
    title = "{Grey Galaxies{\textquoteright} as an endpoint of the Kerr-AdS superradiant instability}",
    eprint = "2305.08922",
    archivePrefix = "arXiv",
    primaryClass = "hep-th",
    doi = "10.1007/JHEP11(2023)024",
    journal = "JHEP",
    volume = "11",
    pages = "024",
    year = "2023"
}

@article{Choi:2024xnv,
    author = "Choi, Sunjin and Jain, Diksha and Kim, Seok and Krishna, Vineeth and Lee, Eunwoo and Minwalla, Shiraz and Patel, Chintan",
    title = "{Dual dressed black holes as the end point of the charged superradiant instability in $\mathcal{N} = 4$ Yang Mills}",
    eprint = "2409.18178",
    archivePrefix = "arXiv",
    primaryClass = "hep-th",
    reportNumber = "TIFR/TH/24-19, LCTP-24-17",
    doi = "10.21468/SciPostPhys.18.4.137",
    journal = "SciPost Phys.",
    volume = "18",
    number = "4",
    pages = "137",
    year = "2025"
}

@article{Fiola:1994ir,
    author = "Fiola, Thomas M. and Preskill, John and Strominger, Andrew and Trivedi, Sandip P.",
    title = "{Black hole thermodynamics and information loss in two-dimensions}",
    eprint = "hep-th/9403137",
    archivePrefix = "arXiv",
    reportNumber = "CALT-68-1918",
    doi = "10.1103/PhysRevD.50.3987",
    journal = "Phys. Rev. D",
    volume = "50",
    pages = "3987--4014",
    year = "1994"
}

@article{Soni:2024oim,
    author = "Soni, Ronak M.",
    title = "{Extremality as a consistency condition on subregion duality}",
    eprint = "2403.19562",
    archivePrefix = "arXiv",
    primaryClass = "hep-th",
    doi = "10.21468/SciPostPhys.17.5.133",
    journal = "SciPost Phys.",
    volume = "17",
    number = "5",
    pages = "133",
    year = "2024"
}

@article{Haehl:2017sot,
    author = "Haehl, Felix M. and Hijano, Eliot and Parrikar, Onkar and Rabideau, Charles",
    title = "{Higher Curvature Gravity from Entanglement in Conformal Field Theories}",
    eprint = "1712.06620",
    archivePrefix = "arXiv",
    primaryClass = "hep-th",
    doi = "10.1103/PhysRevLett.120.201602",
    journal = "Phys. Rev. Lett.",
    volume = "120",
    number = "20",
    pages = "201602",
    year = "2018"
}

@article{Levine:2020upy,
    author = "Levine, Adam and Shahbazi-Moghaddam, Arvin and Soni, Ronak M.",
    title = "{Seeing the entanglement wedge}",
    eprint = "2009.11305",
    archivePrefix = "arXiv",
    primaryClass = "hep-th",
    doi = "10.1007/JHEP06(2021)134",
    journal = "JHEP",
    volume = "06",
    pages = "134",
    year = "2021"
}

@article{Balasubramanian:2018axm,
    author = "Balasubramanian, Vijay and Parrikar, Onkar",
    title = "{Remarks on entanglement entropy in string theory}",
    eprint = "1801.03517",
    archivePrefix = "arXiv",
    primaryClass = "hep-th",
    doi = "10.1103/PhysRevD.97.066025",
    journal = "Phys. Rev. D",
    volume = "97",
    number = "6",
    pages = "066025",
    year = "2018"
}

@article{Belin:2018bpg,
    author = "Belin, Alexandre and Lewkowycz, Aitor and S{\'a}rosi, G{\'a}bor",
    title = "{Complexity and the bulk volume, a new York time story}",
    eprint = "1811.03097",
    archivePrefix = "arXiv",
    primaryClass = "hep-th",
    doi = "10.1007/JHEP03(2019)044",
    journal = "JHEP",
    volume = "03",
    pages = "044",
    year = "2019"
}

@article{Haehl:2019fjz,
    author = "Haehl, Felix M. and Mintun, Eric and Pollack, Jason and Speranza, Antony J. and Van Raamsdonk, Mark",
    title = "{Nonlocal multi-trace sources and bulk entanglement in holographic conformal field theories}",
    eprint = "1904.01584",
    archivePrefix = "arXiv",
    primaryClass = "hep-th",
    doi = "10.1007/JHEP06(2019)005",
    journal = "JHEP",
    volume = "06",
    pages = "005",
    year = "2019"
}

@article{Dong:2017xht,
    author = "Dong, Xi and Lewkowycz, Aitor",
    title = "{Entropy, Extremality, Euclidean Variations, and the Equations of Motion}",
    eprint = "1705.08453",
    archivePrefix = "arXiv",
    primaryClass = "hep-th",
    doi = "10.1007/JHEP01(2018)081",
    journal = "JHEP",
    volume = "01",
    pages = "081",
    year = "2018"
}

@article{Wall:2011hj,
    author = "Wall, Aron C.",
    title = "{A proof of the generalized second law for rapidly changing fields and arbitrary horizon slices}",
    eprint = "1105.3445",
    archivePrefix = "arXiv",
    primaryClass = "gr-qc",
    doi = "10.1103/PhysRevD.85.104049",
    journal = "Phys. Rev. D",
    volume = "85",
    pages = "104049",
    year = "2012",
    note = "[Erratum: Phys.Rev.D 87, 069904 (2013)]"
}

@article{Flanagan:1996gw,
    author = "Flanagan, Eanna E. and Wald, Robert M.",
    title = "{Does back reaction enforce the averaged null energy condition in semiclassical gravity?}",
    eprint = "gr-qc/9602052",
    archivePrefix = "arXiv",
    reportNumber = "EFI-96-08",
    doi = "10.1103/PhysRevD.54.6233",
    journal = "Phys. Rev. D",
    volume = "54",
    pages = "6233--6283",
    year = "1996"
}

@article{Parrikar:2024zbb,
    author = "Parrikar, Onkar and Rajgadia, Harshit and Singh, Vivek and Sorce, Jonathan",
    title = "{Relational bulk reconstruction from modular flow}",
    eprint = "2403.02377",
    archivePrefix = "arXiv",
    primaryClass = "hep-th",
    reportNumber = "MIT-CTP/5688",
    doi = "10.1007/JHEP07(2024)138",
    journal = "JHEP",
    volume = "07",
    pages = "138",
    year = "2024"
}

@article{Lashkari:2015hha,
    author = "Lashkari, Nima and Van Raamsdonk, Mark",
    title = "{Canonical Energy is Quantum Fisher Information}",
    eprint = "1508.00897",
    archivePrefix = "arXiv",
    primaryClass = "hep-th",
    doi = "10.1007/JHEP04(2016)153",
    journal = "JHEP",
    volume = "04",
    pages = "153",
    year = "2016"
}

@article{Faulkner:2013ica,
    author = "Faulkner, Thomas and Guica, Monica and Hartman, Thomas and Myers, Robert C. and Van Raamsdonk, Mark",
    title = "{Gravitation from Entanglement in Holographic CFTs}",
    eprint = "1312.7856",
    archivePrefix = "arXiv",
    primaryClass = "hep-th",
    doi = "10.1007/JHEP03(2014)051",
    journal = "JHEP",
    volume = "03",
    pages = "051",
    year = "2014"
}

@article{Jafferis:2015del,
    author = "Jafferis, Daniel L. and Lewkowycz, Aitor and Maldacena, Juan and Suh, S. Josephine",
    title = "{Relative entropy equals bulk relative entropy}",
    eprint = "1512.06431",
    archivePrefix = "arXiv",
    primaryClass = "hep-th",
    reportNumber = "NSF-KITP-15-162",
    doi = "10.1007/JHEP06(2016)004",
    journal = "JHEP",
    volume = "06",
    pages = "004",
    year = "2016"
}

@article{Faulkner:2017tkh,
    author = "Faulkner, Thomas and Haehl, Felix M. and Hijano, Eliot and Parrikar, Onkar and Rabideau, Charles and Van Raamsdonk, Mark",
    title = "{Nonlinear Gravity from Entanglement in Conformal Field Theories}",
    eprint = "1705.03026",
    archivePrefix = "arXiv",
    primaryClass = "hep-th",
    doi = "10.1007/JHEP08(2017)057",
    journal = "JHEP",
    volume = "08",
    pages = "057",
    year = "2017"
}

@article{Lee:1990nz,
    author = "Lee, J. and Wald, Robert M.",
    title = "{Local symmetries and constraints}",
    doi = "10.1063/1.528801",
    journal = "J. Math. Phys.",
    volume = "31",
    pages = "725--743",
    year = "1990"
}

@article{Wald:1993nt,
    author = "Wald, Robert M.",
    title = "{Black hole entropy is the Noether charge}",
    eprint = "gr-qc/9307038",
    archivePrefix = "arXiv",
    reportNumber = "EFI-93-42",
    doi = "10.1103/PhysRevD.48.R3427",
    journal = "Phys. Rev. D",
    volume = "48",
    number = "8",
    pages = "R3427--R3431",
    year = "1993"
}

@article{Iyer:1994ys,
    author = "Iyer, Vivek and Wald, Robert M.",
    title = "{Some properties of Noether charge and a proposal for dynamical black hole entropy}",
    eprint = "gr-qc/9403028",
    archivePrefix = "arXiv",
    doi = "10.1103/PhysRevD.50.846",
    journal = "Phys. Rev. D",
    volume = "50",
    pages = "846--864",
    year = "1994"
}

@article{Iyer:1995kg,
    author = "Iyer, Vivek and Wald, Robert M.",
    title = "{A Comparison of Noether charge and Euclidean methods for computing the entropy of stationary black holes}",
    eprint = "gr-qc/9503052",
    archivePrefix = "arXiv",
    doi = "10.1103/PhysRevD.52.4430",
    journal = "Phys. Rev. D",
    volume = "52",
    pages = "4430--4439",
    year = "1995"
}

@article{Wald:1999wa,
    author = "Wald, Robert M. and Zoupas, Andreas",
    title = "{A General definition of 'conserved quantities' in general relativity and other theories of gravity}",
    eprint = "gr-qc/9911095",
    archivePrefix = "arXiv",
    doi = "10.1103/PhysRevD.61.084027",
    journal = "Phys. Rev. D",
    volume = "61",
    pages = "084027",
    year = "2000"
}

@article{Hollands:2012sf,
    author = "Hollands, Stefan and Wald, Robert M.",
    title = "{Stability of Black Holes and Black Branes}",
    eprint = "1201.0463",
    archivePrefix = "arXiv",
    primaryClass = "gr-qc",
    doi = "10.1007/s00220-012-1638-1",
    journal = "Commun. Math. Phys.",
    volume = "321",
    pages = "629--680",
    year = "2013"
}

@article{Harlow:2019yfa,
    author = "Harlow, Daniel and Wu, Jie-Qiang",
    title = "{Covariant phase space with boundaries}",
    eprint = "1906.08616",
    archivePrefix = "arXiv",
    primaryClass = "hep-th",
    doi = "10.1007/JHEP10(2020)146",
    journal = "JHEP",
    volume = "10",
    pages = "146",
    year = "2020"
}

@article{Faulkner:2013ana,
    author = "Faulkner, Thomas and Lewkowycz, Aitor and Maldacena, Juan",
    title = "{Quantum corrections to holographic entanglement entropy}",
    eprint = "1307.2892",
    archivePrefix = "arXiv",
    primaryClass = "hep-th",
    doi = "10.1007/JHEP11(2013)074",
    journal = "JHEP",
    volume = "11",
    pages = "074",
    year = "2013"
}

@article{Lewkowycz:2018sgn,
    author = "Lewkowycz, Aitor and Parrikar, Onkar",
    title = "{The holographic shape of entanglement and Einstein\textquoteright{}s equations}",
    eprint = "1802.10103",
    archivePrefix = "arXiv",
    primaryClass = "hep-th",
    doi = "10.1007/JHEP05(2018)147",
    journal = "JHEP",
    volume = "05",
    pages = "147",
    year = "2018"
}

@article{Lashkari:2019ixo,
    author = "Lashkari, Nima",
    title = "{Modular zero modes and sewing the states of QFT}",
    eprint = "1911.11153",
    archivePrefix = "arXiv",
    primaryClass = "hep-th",
    doi = "10.1007/JHEP04(2021)189",
    journal = "JHEP",
    volume = "21",
    pages = "189",
    year = "2020"
}

@article{crnkovic1987covariant,
  title={Covariant description of canonical formalism in geometrical theories.},
  author={Crnkovic, Cedomir and Witten, Edward},
  journal={Three hundred years of gravitation},
  pages={676--684},
  year={1987}
}

@article{Gibbons:1976ue,
    author = "Gibbons, G. W. and Hawking, S. W.",
    title = "{Action Integrals and Partition Functions in Quantum Gravity}",
    reportNumber = "PRINT-76-0995 (CAMBRIDGE)",
    doi = "10.1103/PhysRevD.15.2752",
    journal = "Phys. Rev. D",
    volume = "15",
    pages = "2752--2756",
    year = "1977"
}

@article{York:1972sj,
    author = "York, Jr., James W.",
    title = "{Role of conformal three geometry in the dynamics of gravitation}",
    doi = "10.1103/PhysRevLett.28.1082",
    journal = "Phys. Rev. Lett.",
    volume = "28",
    pages = "1082--1085",
    year = "1972"
}

@article{Barnich:2007bf,
    author = "Barnich, Glenn and Compere, Geoffrey",
    title = "{Surface charge algebra in gauge theories and thermodynamic integrability}",
    eprint = "0708.2378",
    archivePrefix = "arXiv",
    primaryClass = "gr-qc",
    reportNumber = "ULB-TH-06-30",
    doi = "10.1063/1.2889721",
    journal = "J. Math. Phys.",
    volume = "49",
    pages = "042901",
    year = "2008"
}

@article{Kirklin:2019ror,
    author = "Kirklin, Josh",
    title = "{The Holographic Dual of the Entanglement Wedge Symplectic Form}",
    eprint = "1910.00457",
    archivePrefix = "arXiv",
    primaryClass = "hep-th",
    doi = "10.1007/JHEP01(2020)071",
    journal = "JHEP",
    volume = "01",
    pages = "071",
    year = "2020"
}

@article{Donnelly:2016auv,
    author = "Donnelly, William and Freidel, Laurent",
    title = "{Local subsystems in gauge theory and gravity}",
    eprint = "1601.04744",
    archivePrefix = "arXiv",
    primaryClass = "hep-th",
    doi = "10.1007/JHEP09(2016)102",
    journal = "JHEP",
    volume = "09",
    pages = "102",
    year = "2016"
}

@article{Brown:1992br,
    author = "Brown, J. David and York, Jr., James W.",
    title = "{Quasilocal energy and conserved charges derived from the gravitational action}",
    eprint = "gr-qc/9209012",
    archivePrefix = "arXiv",
    reportNumber = "IFP-423-UNC, TAR-009-UNC",
    doi = "10.1103/PhysRevD.47.1407",
    journal = "Phys. Rev. D",
    volume = "47",
    pages = "1407--1419",
    year = "1993"
}

@article{Balasubramanian:1999re,
    author = "Balasubramanian, Vijay and Kraus, Per",
    title = "{A Stress tensor for Anti-de Sitter gravity}",
    eprint = "hep-th/9902121",
    archivePrefix = "arXiv",
    reportNumber = "HUTP-99-A002, EFI-99-6, NSF-ITP-98-132",
    doi = "10.1007/s002200050764",
    journal = "Commun. Math. Phys.",
    volume = "208",
    pages = "413--428",
    year = "1999"
}

@book{wald2010general,
  title={General relativity},
  author={Wald, Robert M},
  year={2010},
  publisher={University of Chicago press}
}

@article{Jensen:2023yxy,
    author = "Jensen, Kristan and Sorce, Jonathan and Speranza, Antony J.",
    title = "{Generalized entropy for general subregions in quantum gravity}",
    eprint = "2306.01837",
    archivePrefix = "arXiv",
    primaryClass = "hep-th",
    doi = "10.1007/JHEP12(2023)020",
    journal = "JHEP",
    volume = "12",
    pages = "020",
    year = "2023"
}

@article{connes1978homogeneity,
  title={Homogeneity of the state space of factors of type III1},
  author={Connes, Alain and St{\o}rmer, Erling},
  journal={Journal of Functional Analysis},
  volume={28},
  number={2},
  pages={187--196},
  year={1978},
  publisher={Elsevier}
}

@article{Engelhardt:2018kcs,
    author = "Engelhardt, Netta and Wall, Aron C.",
    title = "{Coarse Graining Holographic Black Holes}",
    eprint = "1806.01281",
    archivePrefix = "arXiv",
    primaryClass = "hep-th",
    doi = "10.1007/JHEP05(2019)160",
    journal = "JHEP",
    volume = "05",
    pages = "160",
    year = "2019"
}

@article{Czech:2023zmq,
    author = "Czech, Bartlomiej and de Boer, Jan and Esp{\'\i}ndola, Ricardo and Najian, Bahman and van der Heijden, Jeremy and Zukowski, Claire",
    title = "{Changing states in holography: From modular Berry curvature to the bulk symplectic form}",
    eprint = "2305.16384",
    archivePrefix = "arXiv",
    primaryClass = "hep-th",
    doi = "10.1103/PhysRevD.108.066003",
    journal = "Phys. Rev. D",
    volume = "108",
    number = "6",
    pages = "066003",
    year = "2023"
}

@article{Czech:2017zfq,
    author = "Czech, Bartlomiej and Lamprou, Lampros and Mccandlish, Samuel and Sully, James",
    title = "{Modular Berry Connection for Entangled Subregions in AdS/CFT}",
    eprint = "1712.07123",
    archivePrefix = "arXiv",
    primaryClass = "hep-th",
    doi = "10.1103/PhysRevLett.120.091601",
    journal = "Phys. Rev. Lett.",
    volume = "120",
    number = "9",
    pages = "091601",
    year = "2018"
}

@article{Czech:2018kvg,
    author = "Czech, Bartlomiej and Lamprou, Lampros and Susskind, Leonard",
    title = "{Entanglement Holonomies}",
    eprint = "1807.04276",
    archivePrefix = "arXiv",
    primaryClass = "hep-th",
    month = "7",
    year = "2018"
}

@article{deBoer:2025rxx,
    author = "de Boer, Jan and Najian, Bahman and van der Heijden, Jeremy and Zukowski, Claire",
    title = "{Modular chaos, operator algebras, and the Berry phase}",
    eprint = "2505.04682",
    archivePrefix = "arXiv",
    primaryClass = "hep-th",
    month = "5",
    year = "2025"
}

@article{Belin:2018fxe,
    author = "Belin, Alexandre and Lewkowycz, Aitor and S{\'a}rosi, G{\'a}bor",
    title = "{The boundary dual of the bulk symplectic form}",
    eprint = "1806.10144",
    archivePrefix = "arXiv",
    primaryClass = "hep-th",
    doi = "10.1016/j.physletb.2018.10.071",
    journal = "Phys. Lett. B",
    volume = "789",
    pages = "71--75",
    year = "2019"
}

@article{Wald:1980jn,
    author = "Wald, Robert M.",
    title = "{Dynamics in nonglobally hyperbolic static space-times}",
    doi = "10.1063/1.524403",
    journal = "J. Math. Phys.",
    volume = "21",
    pages = "2802--2805",
    year = "1980"
}

@article{Ishibashi:2004wx,
    author = "Ishibashi, Akihiro and Wald, Robert M.",
    title = "{Dynamics in nonglobally hyperbolic static space-times. 3. Anti-de Sitter space-time}",
    eprint = "hep-th/0402184",
    archivePrefix = "arXiv",
    doi = "10.1088/0264-9381/21/12/012",
    journal = "Class. Quant. Grav.",
    volume = "21",
    pages = "2981--3014",
    year = "2004"
}

@article{Ishibashi:2003jd,
    author = "Ishibashi, Akihiro and Wald, Robert M.",
    title = "{Dynamics in nonglobally hyperbolic static space-times. 2. General analysis of prescriptions for dynamics}",
    eprint = "gr-qc/0305012",
    archivePrefix = "arXiv",
    doi = "10.1088/0264-9381/20/16/318",
    journal = "Class. Quant. Grav.",
    volume = "20",
    pages = "3815--3826",
    year = "2003"
}

@article{Breitenlohner:1982bm,
    author = "Breitenlohner, Peter and Freedman, Daniel Z.",
    title = "{Positive Energy in anti-De Sitter Backgrounds and Gauged Extended Supergravity}",
    reportNumber = "PRINT-82-0420 (MIT)",
    doi = "10.1016/0370-2693(82)90643-8",
    journal = "Phys. Lett. B",
    volume = "115",
    pages = "197--201",
    year = "1982"
}

@article{Lashkari:2016idm,
    author = "Lashkari, Nima and Lin, Jennifer and Ooguri, Hirosi and Stoica, Bogdan and Van Raamsdonk, Mark",
    title = "{Gravitational positive energy theorems from information inequalities}",
    eprint = "1605.01075",
    archivePrefix = "arXiv",
    primaryClass = "hep-th",
    doi = "10.1093/ptep/ptw139",
    journal = "PTEP",
    volume = "2016",
    number = "12",
    pages = "12C109",
    year = "2016"
}

@article{Hollands:2024vbe,
    author = "Hollands, Stefan and Wald, Robert M. and Zhang, Victor G.",
    title = "{Entropy of dynamical black holes}",
    eprint = "2402.00818",
    archivePrefix = "arXiv",
    primaryClass = "hep-th",
    doi = "10.1103/PhysRevD.110.024070",
    journal = "Phys. Rev. D",
    volume = "110",
    number = "2",
    pages = "024070",
    year = "2024"
}

@article{Wall:2015raa,
    author = "Wall, Aron C.",
    title = "{A Second Law for Higher Curvature Gravity}",
    eprint = "1504.08040",
    archivePrefix = "arXiv",
    primaryClass = "gr-qc",
    doi = "10.1142/S0218271815440149",
    journal = "Int. J. Mod. Phys. D",
    volume = "24",
    number = "12",
    pages = "1544014",
    year = "2015"
}

@article{Chandrasekaran:2021hxc,
    author = "Chandrasekaran, Venkatesa and Flanagan, Eanna E. and Shehzad, Ibrahim and Speranza, Antony J.",
    title = "{Brown-York charges at null boundaries}",
    eprint = "2109.11567",
    archivePrefix = "arXiv",
    primaryClass = "hep-th",
    doi = "10.1007/JHEP01(2022)029",
    journal = "JHEP",
    volume = "01",
    pages = "029",
    year = "2022"
}

@article{Chandrasekaran:2021vyu,
    author = "Chandrasekaran, Venkatesa and Flanagan, Eanna E. and Shehzad, Ibrahim and Speranza, Antony J.",
    title = "{A general framework for gravitational charges and holographic renormalization}",
    eprint = "2111.11974",
    archivePrefix = "arXiv",
    primaryClass = "gr-qc",
    doi = "10.1142/S0217751X22501056",
    journal = "Int. J. Mod. Phys. A",
    volume = "37",
    number = "17",
    pages = "2250105",
    year = "2022"
}

@article{Flanagan:2015pxa,
    author = "Flanagan, {\'E}anna {\'E}. and Nichols, David A.",
    title = "{Conserved charges of the extended Bondi-Metzner-Sachs algebra}",
    eprint = "1510.03386",
    archivePrefix = "arXiv",
    primaryClass = "hep-th",
    doi = "10.1103/PhysRevD.95.044002",
    journal = "Phys. Rev. D",
    volume = "95",
    number = "4",
    pages = "044002",
    year = "2017",
    note = "[Erratum: Phys.Rev.D 108, 069902 (2023)]"
}

@article{Hawking:2016msc,
    author = "Hawking, Stephen W. and Perry, Malcolm J. and Strominger, Andrew",
    title = "{Soft Hair on Black Holes}",
    eprint = "1601.00921",
    archivePrefix = "arXiv",
    primaryClass = "hep-th",
    doi = "10.1103/PhysRevLett.116.231301",
    journal = "Phys. Rev. Lett.",
    volume = "116",
    number = "23",
    pages = "231301",
    year = "2016"
}

@article{Hawking:2016sgy,
    author = "Hawking, Stephen W. and Perry, Malcolm J. and Strominger, Andrew",
    title = "{Superrotation Charge and Supertranslation Hair on Black Holes}",
    eprint = "1611.09175",
    archivePrefix = "arXiv",
    primaryClass = "hep-th",
    doi = "10.1007/JHEP05(2017)161",
    journal = "JHEP",
    volume = "05",
    pages = "161",
    year = "2017"
}

@article{Haco:2018ske,
    author = "Haco, Sasha and Hawking, Stephen W. and Perry, Malcolm J. and Strominger, Andrew",
    title = "{Black Hole Entropy and Soft Hair}",
    eprint = "1810.01847",
    archivePrefix = "arXiv",
    primaryClass = "hep-th",
    doi = "10.1007/JHEP12(2018)098",
    journal = "JHEP",
    volume = "12",
    pages = "098",
    year = "2018"
}

@article{He:2020ifr,
    author = "He, Temple and Mitra, Prahar",
    title = "{Covariant Phase Space and Soft Factorization in Non-Abelian Gauge Theories}",
    eprint = "2009.14334",
    archivePrefix = "arXiv",
    primaryClass = "hep-th",
    doi = "10.1007/JHEP03(2021)015",
    journal = "JHEP",
    volume = "03",
    pages = "015",
    year = "2021"
}

@article{He:2023bvv,
    author = "He, Temple and Mitra, Prahar",
    title = "{Asymptotic structure of higher dimensional Yang-Mills theory}",
    eprint = "2306.04571",
    archivePrefix = "arXiv",
    primaryClass = "hep-th",
    reportNumber = "CALT-TH 2023-015",
    doi = "10.21468/SciPostPhys.16.5.142",
    journal = "SciPost Phys.",
    volume = "16",
    number = "5",
    pages = "142",
    year = "2024"
}

@article{Ciambelli:2021nmv,
    author = "Ciambelli, Luca and Leigh, Robert G. and Pai, Pin-Chun",
    title = "{Embeddings and Integrable Charges for Extended Corner Symmetry}",
    eprint = "2111.13181",
    archivePrefix = "arXiv",
    primaryClass = "hep-th",
    doi = "10.1103/PhysRevLett.128.171302",
    journal = "Phys. Rev. Lett.",
    volume = "128",
    year = "2022"
}

@article{Ciambelli:2022cfr,
    author = "Ciambelli, Luca and Leigh, Robert G.",
    title = "{Universal corner symmetry and the orbit method for gravity}",
    eprint = "2207.06441",
    archivePrefix = "arXiv",
    primaryClass = "hep-th",
    doi = "10.1016/j.nuclphysb.2022.116053",
    journal = "Nucl. Phys. B",
    volume = "986",
    pages = "116053",
    year = "2023"
}

@article{Ciambelli:2022vot,
    author = "Ciambelli, Luca",
    title = "{From Asymptotic Symmetries to the Corner Proposal}",
    eprint = "2212.13644",
    archivePrefix = "arXiv",
    primaryClass = "hep-th",
    doi = "10.22323/1.435.0002",
    journal = "PoS",
    volume = "Modave2022",
    pages = "002",
    year = "2023"
}

@article{Hartle:1981zt,
    author = "Hartle, J. B. and Horowitz, G. T.",
    title = "{Ground State Expectation Value of the Metric in the 1/$N$ or Semiclassical Approximation to Quantum Gravity}",
    doi = "10.1103/PhysRevD.24.257",
    journal = "Phys. Rev. D",
    volume = "24",
    pages = "257--274",
    year = "1981"
}
\bibliographystyle{JHEP}

\end{document}